\documentclass[fleqn,usenatbib,useAMS]{mnras}
\usepackage{times}
\usepackage{mathptmx}
\usepackage[T1]{fontenc}
\usepackage{ae,aecompl}
\usepackage{graphicx}	
\usepackage{amsmath}	
\usepackage{amssymb}	

\newcommand{\Msun}{\ensuremath{\,{\rm M}_\odot}}                  
\newcommand{\Rsun}{\ensuremath{\,{\rm R}_\odot}}                  
\newcommand{\Teff}{\ensuremath{T_{\rm eff}}}                      
\newcommand{\Mjup}{\ensuremath{\,{\rm M}_{\rm Jup}}}              
\newcommand{\Rjup}{\ensuremath{\,{\rm R}_{\rm Jup}}}              
\newcommand{\Teq}{\ensuremath{T_{\rm eq}^{\,\prime}}}             
\newcommand{\safronov}{\ensuremath{\Theta}}                       
\newcommand{\kms}{\,km\,s$^{-1}$}                                 
\newcommand{\ms}{\,m\,s$^{-1}$}                                   
\newcommand{\mss}{\,m\,s$^{-2}$}                                  
\newcommand{\as}{\ensuremath{^{\prime\prime}}}                    
\newcommand{\am}{\ensuremath{^\prime}}                            
\newcommand{\FeH}{\ensuremath{\left[\frac{\rm Fe}{\rm H}\right]}} 
\newcommand{\pjup}{\ensuremath{\,\rho_{\rm Jup}}}                 
\newcommand{\psun}{\ensuremath{\,\rho_\odot}}                     

\newcommand{\mcc}[1]{\multicolumn{3}{c}{#1}}

\newcommand{\ermcc}[5]{\mcc{\ensuremath{{#1\,^{+#2}_{-#3}}\,^{+#4}_{-#5}}}}

\newcommand{\prism}{\textsc{prism}}	
\newcommand{\gemc}{\textsc{gemc}}	
\newcommand{\jktebop}{\textsc{jktebop}}	

\title[Possible bimodal cloud distribution in HAT-P-32\,A\,b]
{Possible detection of a bimodal cloud distribution in the atmosphere of HAT-P-32\,A\,b from multi-band photometry}

\author[Tregloan-Reed et al.]
       {Jeremy Tregloan-Reed\,$^{1}$\thanks{Email: j.j.tregloan.reed@gmail.com},
        John Southworth\,$^{2}$,
        L.\ Mancini\,$^{3,\,4,\,5}$,
       P. Molli\`ere\,$^{4,\,6}$,
 \newauthor
        S.\ Ciceri\,$^{7}$,
        I.\ Bruni\,$^{5}$,
        D.\ Ricci\,$^{8,\,9}$,
        C.\ Ayala-Loera\,$^{8,\,10}$,
        T.\ Henning\,$^{4}$
        \\
        \\
        $^{1}$\,Carl Sagan Center, SETI Institute, Mountain View, CA 94043, USA \\
        $^{2}$\,Astrophysics Group, Keele University, Staffordshire, ST5 5BG, UK \\
        $^{3}$\,Department of Physics, University of Rome Tor Vergata, Via della Ricerca Scientifica 1, 00133 -- Roma, Italy \\
        $^{4}$\,Max Planck Institute for Astronomy, K\"onigstuhl 17, 69117 Heidelberg, Germany \\
        $^{5}$\,INAF--Osservatorio Astrofsico di Torino, via Osservatorio 20, 10025--Pino Torinese, Italy \\
        $^{6}$\,Leiden Observatory, Leiden University, Postbus 9513, 2300 RA Leiden, The Netherlands \\
        $^{7}$\,Department of Astronomy, Stockholm University, 11419, Stockholm, Sweden \\
        $^{8}$\,Observatorio Astron\'omico Nacional, Instituto de Astronom\'ia --Universidad Nacional Aut\'onoma de M\'exico, Ap. P. 877, Ensenada, \\  \, BC 22860, Mexico \\
        $^{9}$\,Instituto de Astrof\'isica de Canarias E-38205 La Laguna, Tenerife, Spain; Universidad de La Laguna, Departmento de Astrof\'isica, \\  \, E-38205, La Laguna, Tenerife, Spain \\
        $^{10}$\,Observat\'orio Nacional/MCTI, Rua Gen. Jos\'e Cristino, 77, 20921-400, Rio de Janeiro, Brazil \\
        \vspace*{-1.1cm}
}

\date{Accepted 2017 November 29. Received 2017 November 20; in original form 2017 July 07}

\pubyear{2017}

\begin{document}
\label{firstpage}
\pagerange{\pageref{firstpage}--\pageref{lastpage}}
\maketitle 

\begin{abstract}
We present high-precision photometry of eight separate transit events in the HAT-P-32 planetary system. One transit event was observed simultaneously by two telescopes of which one obtained a simultaneous multi-band light curve in three optical bands, giving a total of 11 transit light curves. Due to the filter selection and in conjunction with using the defocussed photometry technique we were able to obtain an extremely high precision, ground-based transit in the \textit{u}-band (350\,nm), with an rms scatter of $\approx 1$\,mmag. All 11 transits were modelled using \prism\ and \gemc, and the physical properties of the system calculated. We find the mass and radius of the host star to be $1.182\pm 0.041\Msun$ and $1.225\pm0.015\Rsun$, respectively. For the planet we find a mass of $0.80\pm 0.14\Mjup$, a radius of $1.807\pm0.022\Rjup$ and a density of $0.126\pm0.023\pjup$. These values are consistent with those found in the literature. We also obtain a new orbital ephemeris for the system $ T_0 = {\rm BJD/TDB} \,\, 2\,454\,420.447187 (96) \, + \,2.15000800 (10) \times E $. We measured the transmission spectrum of HAT-P-32\,A\,b and compared it to theoretical transmission spectra. Our results indicate a bimodal cloud particle distribution consisting of Rayleigh--like haze and grey absorbing cloud particles within the atmosphere of HAT-P-32\,A\,b.
\end{abstract}



\section{Introduction}
\label{sec:intro}

The number of currently known extrasolar planets exceeds 3600\footnote{See \href{http://exoplanet.eu} {\tt http://exoplanet.eu} \citep{ExoWebPaper}.}, while, the total number of known transiting extrasolar planets (TEPs) exceeds 1400\footnote{See TEPCat \citep[Transiting Extrasolar Planet Catalogue;][]{Me11mn} at: \href{http://www.astro.keele.ac.uk/jkt/tepcat/}{\tt http://www.astro.keele.ac.uk/jkt/tepcat/}.}. The majority of TEPs have been discovered from ground-based (e.g.,\ SuperWasp: \citealt{SuperWasp}; HATNet: \citealt{HAT}) or space-based (CoRoT: \citealt{Corot}; \textit{Kepler}: \citealt{Kepler}) transit surveys, and later confirmed by use of the radial velocity technique \citep{Butler1996,Butler1999,Queloz2000}. The majority of these are small objects discovered by \textit{Kepler}, however, there are difficulties in studying these objects due to their small size and their long orbital periods.

With the development of the \textit{NGTS} planet hunter \citep{NGTS} and the NASA \textit{TESS} satellite \citep{TESS} we are entering a new era of planetary transit detection. These new surveys are expected to find both mini-Neptune and rocky planets orbiting K-dwarf and M-dwarf stars within our solar neighbourhood, which will be suitable for ground based follow-up observations, especially those aimed to probe planetary atmospheres. This is a key step to finding an Earth-like planet elsewhere in the galaxy, as it will allow for detailed atmospheric studies. At present, due to observational constraints, the majority of TEPs suitable for in-depth studies are hot Jupiters (e.g., WASP-19\,b: \citealt{Hellier2011}; WASP-12\,b: \citealt{Sing2013}; WASP-31\,b: \citealt{Sing2015}; HAT-P-1\,b: \citealt{Nikolov2014}; WASP-6\,b: \citealt{Nikolov2015}; WASP-39\,b: \citealt{Nikolov2016, Fischer2016}; WASP-98\,b: \citealt{Mancini2016}).

Transit spectroscopy can be used to study an exoplanet's atmosphere \citep{Seager2000,Charbonneau2002}, where measurements of the planetary radius can be made for different wavelengths. The results are then compared to theoretical model atmospheres \citep[e.g.,][]{Irwin2008,Fortney2008,Madhusudhan2009}, to determine the chemical composition of the outer planetary atmosphere. However, this can be hampered by condensates that can weaken or mask spectral features depending on the height of the cloud deck \citep[e.g.,][]{Sudarsky2003,Fortney2005}. Some theoretical models predict the presence of spectrally active atmospheric constituents such as TiO and VO, which have been considered responsible for causing temperature inversions \citep{Hubeny2003, Fortney2008, Fortney2010, Burrows2010}. These spectral signatures can be observed in the optical UV--blue region ($\approx$\,$350$--$450$\,nm). Observations have been made in the optical UV--blue using transit spectroscopy \citep[e.g.,][]{Sing2013} and have discovered an increase in the planetary radius towards bluer wavelengths, indicative of a Rayleigh scattering slope (e.g.,  GJ\,3470\,b: \citealt{Dragomir2015}; WASP-31\,b: \citealt{Sing2015}; HAT-P-1\,b: \citealt{Nikolov2014}; WASP-6\,b: \citealt{Nikolov2015}; WASP-39\,b: \citealt{Nikolov2016, Fischer2016}).

For highly irradiated planets, the atmosphere at optical wavelengths is a vital part of the energy budget of the planet, as it is where the bulk of the stellar flux is deposited \citep{Sing2011}. By using multi-band imagers \citep[e.g., GROND, on the MPG 2.2\,m telescope, ESO La Silla, Chile,][]{Greiner2008} it is possible to view a transit simultaneously in multiple wavelengths. This then allows variations as small as an atmospheric scale height in the planetary radius to be observed over the filter FWHM (for a Cousins R filter, FWHM $=158$\,nm). Such variations can arise from Rayleigh scattering, Mie scattering and from molecular opacities, so are tracers of the atmospheric conditions and chemical composition \citep[e.g.,][]{Sou2012,Sou2015,Mancini2013a,Mancini2013b,Mancini2014, Chen2014}. By using a wide wavelength range a broadband transmission spectrum can be constructed \citep[e.g.,][]{Nikolov2013}.

One of the inherent difficulties in using ground-based simultaneous multi-band defocused photometry lies in the fact that the amount of defocussing is optimised for a single filter (for optimal precision this is normally an \textit{r}-filter). Subsequently the quality of the transit data reduces for the other filters. This usually results in a poor quality light curve in the \textit{u}-band (e.g., HAT-P-5: \citealt{Sou2012}; WASP-57: \citealt{Sou2015}; HAT-P-8: \citealt{Mancini2013}; HAT-P-23; WASP-48: \citealt{Ciceri2015}) and is unsuitable for use in the comparison between measured planetary radii and theoretical atmospheric predictions. It also hinders the detection of a possible near-UV Rayleigh scattering slope.


\subsection{Previous work on HAT-P-32}

The transiting planetary system HAT-P-32 was discovered by \citet{Hartman2011} using photometry from the HATNet telescope. They determined an orbital period of $P = 2.15$\,days for the planet HAT-P-32\,A\,b. Reconnaissance spectroscopy and RV measurements were obtained using the Harvard-Smithsonian Center for Astrophysics (CfA) Digital Speedometer (DS; \citealt{Latham1992}) on the FLWO 1.5m telescope. \citet{Hartman2011} determined for a circular orbit that the stellar mass and radius are $M_\star = 1.160\pm0.041$\Msun\ and $R_\star = 1.219\pm0.016$\Rsun, respectively. They found the planetary mass and radius to be $M_{\rm p} = 0.860\pm0.164$\Mjup\ and $R_{\rm p} = 1.789\pm0.025$\Rjup. They mentioned difficulties in precisely determining the stellar and planetary properties due to high velocity jitter ($\approx 80$\ms). From the spectroscopic data they determined a value for the projected stellar rotational velocity (for a circular orbit) of $v\sin I = 20.7 \pm 0.5$\kms\ and a macroturbulence ($v_{\rm mac}$) value of $4.69$\kms.

Between 2008 and 2011 \citet{Sada2012} used the FLAMINGOS infrared camera\footnote{The observations were preformed in the J-, H- and JH-bands.} on the 2.1\,m Kitt Peak National Observatory Telescope to observe 57 transits of 32 known exoplanets, with the HAT-P-32 planetary system being one of the targets. They observed two separate transits, with one observed simultaneously with two telescopes. With the data \citet{Sada2012} were able to further refine the orbital ephemeris, orbital inclination, ratio of the radii and the scaled stellar radius.

Between 2012 and 2014 \citet{Adams2012,Adams2013,Dressing2014} conducted an exhaustive adaptive optics imagining campaign of 15 known TEPs and 189 \textit{Kepler} Objects of Interests (KOIs). During this campaign they observed HAT-P-32\,A and discovered a faint companion, HAT-P-32\,B at a distance of 2.9\as\ combined with a magnitude difference of $\Delta K_s = 3.4$ \citep{Adams2013}, which was just beyond the detection limit of \citet{Hartman2011}.

The atmosphere of HAT-P-32\,A\,b was studied via transit spectroscopy \citep{Gibson2013}, using GMOS on the Gemini North telescope. Two transits were observed and, using differential spectro-photometry, a white light curve and 29 spectral light curves were generated for each transit. From this \citet{Gibson2013} were able to produce a transmission spectrum of the atmosphere of HAT-P-32\,A\,b covering 520--930\,nm. From their work \citet{Gibson2013} was able to refine the system parameters further and found the orbital inclination to be $89.12^{+0.61}_{-0.68}$ degrees, and the planetary radius and density to be $R_{\rm p} = 1.796^{+0.028}_{-0.027}$\Rjup\ and $\rho_{\rm p} = 0.18\pm0.04$\pjup\ respectively. The examination of the transmission spectrum revealed a flat-spectrum devoid of any broad features larger than one atmospheric scale height. \citet{Gibson2013} concluded that clouds in the upper-atmosphere were acting as a grey absorber.

\citet{Seeliger2014} performed a Transit Timing Variation (TTV) analysis of the HAT-P-32 planetary system to determine the presence of a second planetary body orbiting HAT-P-32\,A. They observed 45 transits by using several telescopes and in particular, telescopes which are part of the YETI\footnote{The Young Exoplanet Transit Initiative \citep{Neuhauser2011}.} network \citep{Seeliger2014}. Using their times of mid-transit and those from the literature, they refined the orbital ephemeris by 21\,ms and found that the data showed no evidence of a TTV signal above 1.5\,min.

\citet{Zhao2014} observed a secondary eclipse of HAT-P-32\,A\,b using \textit{Spitzer}/IRAC at 3.6 and 4.5\,$\mu$m, and Hale/WIRC in the $H$ and $K_s$ bands. Adaptive optics imaging was performed and HAT-P-32\,A and HAT-P-32\,B were visually resolved. The flux ratios of the binary components were measured in six bands (including $r'$ \& $K_s$) and the effective temperature of HAT-P-32\,B was found to be $T_{eff} = 3564 \pm 82$\,K, indicating that HAT-P-32\,B is a M1.5 dwarf star \citep{Zhao2014}. Due to obtaining secondary eclipse timing offset data, \citet{Zhao2014} were able to confirm an orbital eccentricity of HAT-P-32\,A\,b to be $e = 0.0072^{+0.0700}_{-0.0064}$, which is consistent with a circular orbit. \citet{Zhao2014} then compared their secondary eclipse depths with theoretical model atmospheres \citep[e.g.,][]{Fortney2008}. Their analysis showed that the data either matched a temperature inversion caused by a high altitude absorber within the atmosphere of HAT-P-32\,A\,b combined with an inefficient heat redistribution from the day-side to the night-side of the planet, or alternatively a blackbody model with $T_p = 2042\pm50$\,K.

More recently in 2016 three studies into the atmosphere of HAT-P-32\,A\,b were conducted \citep{Mallonn2016a,Mallonn2016b,Nortmann2016}. These studies utilised transit spectroscopy using the Large Binocular Telescope \citep{Mallonn2016a} and the 10.4\,m GTC \citep{Nortmann2016}. The third study used transit photometry from 21 new transit light curves combined with 36 previously published light curves to examine changes in the planetary radius from the near-UV to the near-IR \citep{Mallonn2016b}. All three studies determined a flat spectrum within the range of 500--1000\,nm indicative of high-altitude clouds. However, \citet{Mallonn2016a} makes a tentative detection of a Rayleigh scattering slope below 550\,nm, while in a second study \citet{Mallonn2016b} determined that the tentative detection is less likely due to discrepancies at the reddest wavelengths between the two data sets.


\section{Observations and data reduction}
\label{Sec:data}

\begin{table*} \centering
\begin{footnotesize}
\setlength{\tabcolsep}{4pt}
\caption{\label{tab:obslogh32} Log of the observations presented for HAT-P-32. $N_{\rm obs}$ is the number of observations. `Moon illum.' and
'Moon dist.' are the fractional illumination of the Moon, and its angular distance from HAT-P-32 in degrees, at the midpoint of the transit.}
\begin{tabular}{lcccccccccccc}\hline
Telescope / & Date of & Start time & End time &$N_{\rm obs}$& $T_{exp}$ & $T_{dead}$  & Filter & Airmass & Moon & Moon & Aperture   & Scatter \\
Instrument     & first obs   &  (UT)    &   (UT)   &             & (s) &  (s) &      &         &illum.& dist.& sizes (px) & (mmag)  \\
\hline
CAHA\,1.23 & 2011/08/24 & 23:22 & 04:43 &  212  & 40$\to$75  & 22  & $R$ & 1.62 $\to$ 1.01 & 0.200 & 59.9 & 18, 30, 50 & 3.54 \\
BUSCA & 2011/08/24 & 23:28 & 04:47 &  129  & 120  & 30  & \textit{u} & 1.67 $\to$ 1.03 & 0.200 & 59.9 & 10, 60, 80 & 1.08 \\
BUSCA & 2011/08/24 & 23:28 & 04:45 &  122 & 120 & 30 & \textit{b} & 1.67 $\to$ 1.03 & 0.200 & 59.9 & 15, 20, 80  & 1.04 \\
BUSCA & 2011/08/24 & 23:28 & 04:57 &  125 & 120  & 30  & \textit{y} & 1.67 $\to$ 1.03 & 0.200 & 59.9  & 15, 20, 80  & 0.97 \\
CAHA\,1.23 & 2011/10/04 & 19:57 & 02:30 &  134  & 80$\to$100  & 94  &  $R$ & 1.90 $\to$ 1.01 & 0.588 & 109.4 & 18, 26, 50 & 0.86 \\
CAHA\,1.23 & 2014/01/11 & 19:44 & 00:28 &  77  & 120$\to$150  & 14  & $V$ & 1.03 $\to$ 2.05 & 0.838 & 38.2 & 32, 42, 70 & 0.65 \\
CAHA\,1.23 & 2014/08/31 & 22:18 & 04:39 &  204  & 100  & 11  & $I$ & 1.83 $\to$ 1.01 & 0.354 & 145.2 & 25, 35, 70 & 1.09 \\
SPM\,0.84   & 2014/09/05 & 06:12 & 11:45 &  218  & 40  & 13  &  $R$ & 1.67 $\to$ 1.04 & 0.806 & 108.7 & 15, 35, 40 & 1.61 \\
CAHA\,1.23 & 2014/10/24 & 18:17 & 00:23 &  153 & 120$\to$130  & 12  & $V$ & 2.09 $\to$ 1.01 & 0.010 & 145.9 & 23, 33, 60 & 0.85 \\
Cassini\,1.5 & 2014/12/21 & 16:47 & 22:32 &  172 & 100   & 21  &  $V$ & 1.20 $\to$ 1.00 & 0.003 & 127.2  & 20, 28, 50  & 0.55 \\
CAHA\,1.23 & 2015/08/25 & 22:55 & 04:12 & 181 & 85$\to$100  & 11  & $I$ & 1.78 $\to$ 1.01 & 0.821 & 115.0 & 25, 35, 45 & 0.66 \\
\hline  \end{tabular} \end{footnotesize} \end{table*}


\subsection{BUSCA observation}

BUSCA is capable of viewing a transit simultaneously in four optical passbands, for which three passbands were chosen: Str\"{o}mgren \textit{u}, \textit{b} \& \textit{y}. The fourth passband was neglected due to the need in using filters with the same optical depth. The only available filters with the same optical depth as the Str\"{o}mgren filters were I-band filters (e.g., the Cousins I filter), however, these images were not used due to the target being saturated in the observed images. All four CCDs on BUSCA have a plate scale of 0.176\,\as pixel$^{-1}$ and a field of view of $12\times12$\,\am, and were operated with $2\times 2$ binning. The instrument was defocussed and the telescope was autoguided throughout the observations. Due to the same transit being observed in a Cousins R filter on the CAHA 1.23\,m it was decided to select filters to give observations in the optical UV--blue spectrum. With known difficulties in obtaining precise light curves in the optical UV from simultaneous multi-band photometry \citep[e.g.,][]{Sou2012,Sou2015} the amount of defocusing used was calibrated in the Str\"{o}mgren \textit{b} passband, to optimise the precision of the light curves from all three Str\"{o}mgren passbands. The resulting light curves (labelled U1, B1 and Y1) proved the strategy to be successful with all three light curves having an rms scatter of $\approx 1$\,mmag per point (see Table\,\ref{tab:obslogh32}). In particular the precision in the resulting \textit{u}-band light curve is a major improvement (rms scatter: 1.08\,mmag) on previous \textit{u}-band light curves from simultaneous multi-band photometry (e.g., rms scatter: 3.46\,mmag: \citealt{Sou2015}; 2.37\,mmag: \citealt{Mancini2013}; 3.55\,mmag \& 2.88\,mmag: \citealt{Ciceri2015}).


\subsection{CAHA 1.23\,m telescope observations}
\label{sec:tele}
Six transits of the HAT-P-32 planetary system were observed using the CAHA 1.23\,m telescope, Calar Alto, Spain. The CCD detector has a plate scale 0.32\,\as pixel$^{-1}$ and a field of view of 21.5\am$\times$21.5\am. Two transits were observed using the Johnson V filter (V1:\,2014/01/11 \& V2:\,2014/10/24), two in the Cousins R filter (R1:\,2011/08/24 \& R2:\,2011/10/04) and two in the Cousins I filter (I1:\,2014/08/31 \& I2:\,2015/08/25). All six transits were observed by defocusing the telescope and the telescope was autoguided throughout the observations. The two Johnson V transits were only partially covered, due to an ephemeris error (V1) and scheduling requirements (V2).

The transit I1 proved to be a poor fit. The initial modelling result disagreed with the 1-$\sigma$ uncertainties from the other 10 transits (e.g., $i=86.56^\circ \pm1.10^\circ$). This anomalous result was duplicated when the transit was fitted using a second transit model: \jktebop\ \citep[see][for more details]{Me08mn}. Because the results from both models agreed within their 1-$\sigma$ uncertainties, we concluded that the problem laid within the data itself. Therefore, we decided to only use this transit for the purpose of measuring the time of minimum light.


\subsection{Cassini telescope observation}

A transit of the HAT-P-32 planetary system was observed on 2014/12/21 using BFOSC on the Cassini 1.5\,m telescope, Loiano Observatory, Italy, using a Johnson V filter (labelled V3). The BFOSC  focal-reducing  imager  has a plate scale 0.58\,\as pixel$^{-1}$. The telescope was defocused to allow exposure times of $100$\,s and the pointing of the telescope was maintained throughout the night using the autoguider. The resulting light curve has the lowest rms scatter per point (0.55\,mmag) of the transit light curves presented in this work.


\subsection{San Pedro M\'artir 0.84\,m telescope observation}

A transit of the HAT-P-32 planetary system was observed on 2014/09/05 using the San Pedro M\'artir (SPM) 0.84\,m telescope, Baja California, Mexico, using a Bessell R filter (labelled R3). The telescope was moderately defocused to allow exposure times of $40$\,s and the pointing of the telescope was maintained throughout the night using the autoguider. The transit light curve was obtained as part of the The San Pedro M\'artir Transit Observations Program \citep{Ricci2015, Ricci2017}.


\subsection{Aperture photometry}

We reduced the data in an identical fashion to \citet{Sou2009,Sou2009b,Sou2014}. Aperture photometry was performed with an {\sc idl} implementation of {\sc daophot} \citep{Stetson1987}, and the aperture sizes were adjusted manually on a reference image to obtain the best rms scatter for the out-of-transit data (see Table\,\ref{tab:obslogh32}). A first order polynomial was then fitted to the outside-transit data whilst simultaneously optimising the weights of the comparison stars. Both master bias, sky flat fields and dome flat fields frames were constructed. However, they were left out of the final data reduction as they had little effect on the final reduced science light curves. The resulting data have scatters ranging from 0.551 to 3.540 mmag per point versus a transit fit using \textsc{prism}. The timestamps from the fits files were converted to BJD/TDB. An observing log is given in Table\,\ref{tab:obslogh32}. 


\section{Updates to PRISM \& GEMC}
\label{sec:prism-gemc}

The analysis of the transit light curves presented in this work was conducted by using \prism\ (Planetary Retrospective Integrated Star-spot Model)\footnote{The latest versions of both \prism\ and \gemc\ are directly available from the author via email.} alongside with the optimisation algorithm \gemc\ (Genetic Evolution Markov Chain) \citep[see][]{Jeremy2012, Jeremy2015}. The codes are written in \textsc{idl}\footnote{For further details see \href{http://www.harrisgeospatial.com/ProductsandTechnology/Software/IDL.aspx}{\tt http://www.harrisgeospatial.com}.} (Interactive Data Language) and were developed to model the transit, limb darkening (LD) and starspots on the stellar disc simultaneously. The LD was implemented using the standard quadratic law. \prism\ uses a pixellation approach to represent the star and planet on a two-dimensional array in Cartesian coordinates. Six parameters are used to model the transit: the ratio between the planetary and stellar radii, the sum of the fractional radii\footnote{Where the fractional stellar and planetary radii are defined as the absolute radii scaled by the semimajor axis ($r_{\rm \star,p} = R_{\rm \star,p}/a$).}, the linear and quadratic LD coefficients, the orbital inclination and the time of mid-transit. Then four parameters are used to model each starspot: the longitude and co-latitude of the centre of the starspot on the stellar surface, the angular size of the starspot and the starspot's contrast (the ratio between the intensity ($I$) of the starspot and the surrounding photosphere, $\rho_{spot} = I_{spot}/I_{photo}$).

\gemc\ was created in conjunction to \prism\ to improve the efficiency of finding a global solution in a complex multivariate parameter space compared to conventional MCMC algorithms \citep{Jeremy2012,Jeremy2015}. \gemc\ is a hybrid between an MCMC and a genetic algorithm\footnote{A genetic algorithm mimics biological processes by spawning successive generations of solutions based on breeding and mutation operators from the previous generation.} and is based on the Differential Evolution Markov Chain (DE-MC) put forward by \citet{Cajo2006}. During the `burn-in' stage \gemc\ runs $N$ chains in parallel and for every generation each chain is perturbed by a $P$ dimensional vector within the parameter search space, where $P$ is the number of parameters being fitted. The perturbation vector is orientated within the parameter space, so that the current generation's best-fitting chain lies at the centre of the potential perturbation space. Once the `burn-in' stage is complete and the position of the global solution has been found, \gemc\ switches to a DE-MC algorithm to determine the parameter uncertainties \citep[see][]{Jeremy2015}.

While none of the transit data presented in this work contain any starspot anomalies, so do not require the use of \prism, \prism\ was used to maintain homogeneity with the first author's previous work (WASP-19: \citealt{Jeremy2012}; WASP-50: \citealt{Jeremy2013}; WASP-6: \citealt{Jeremy2015}).

To help facilitate this work, two modifications were made to \prism\ to aid in the analysis of the HAT-P-32 planetary system light curves. To take into account the blended M-dwarf companion, HAT-P-32\,B, a third light ratio parameter was added. A Gaussian prior is used in fitting the third light ratio, to limit the sampled solutions to a Gaussian probability distribution centred around a known flux ratio. The flux ratios used in this work and how they were calculated are given in Section\,\ref{sec:thirdlight}.

The second modification was to add the ability to model and fit the detrending polynomial coefficients used in the detrending of transit data. This is achieved by calculating a new flux value for each model point ($F_i$), by adding an $M^{th}$\,order polynomial (evaluated at the model point) to the original flux ($F_0$) of the model point:
\begin{equation}
F_i = F_0 + \sum_{n=1}^{M} c_n\left( x_i - x_p \right)^{n}
\end{equation}
where $x_p$ is the selected pivot point, $x_i$ is the model points and $c_n$ is the corresponding $n^{th}$ order coefficient. For a data set which has already been detrended by an $M^{th}$\,order polynomial, the optimal solutions for the $M$ detrending coefficients will be zero (e.g., $c_n = 0$), and so, there will be no net change in flux (i.e., $F_i = F_0$).


\section{Data analysis}
\label{Sec:results}

\begin{figure} \includegraphics[width=0.48\textwidth,angle=0]{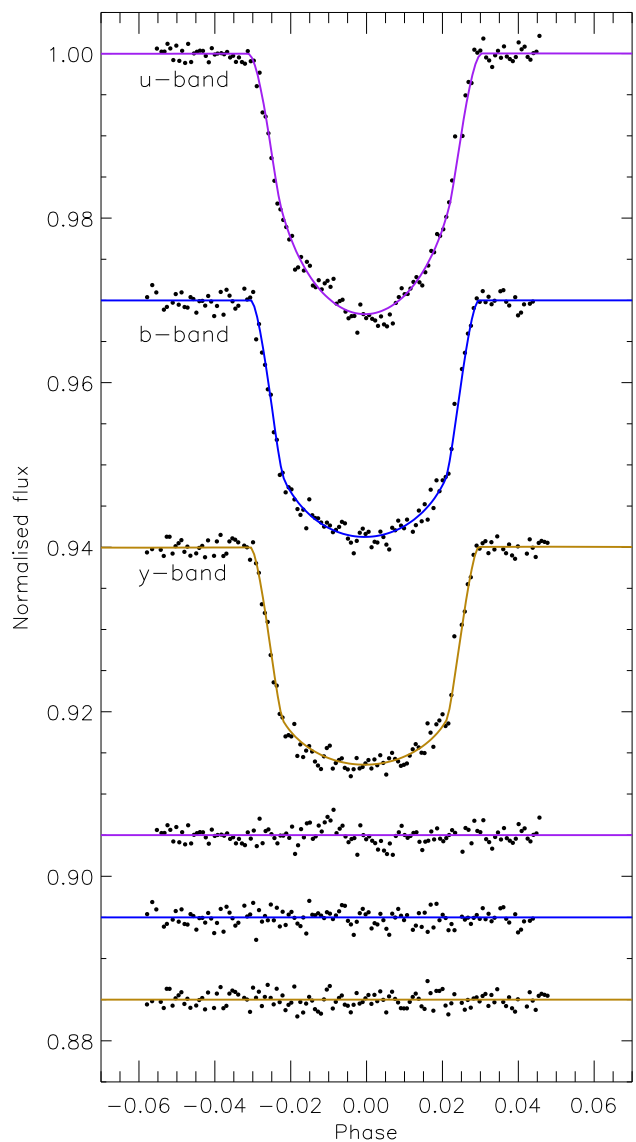}
\caption{\label{fig:busca-lc} Transit light curves, best-fitting models and the residuals of HAT-P-32 from BUSCA. The best fits are shown where purple, blue and gold represents the Str\"omgren \textit{u}, \textit{b} and \textit{y}-bands respectively. The residuals are displayed at the base of the figure.} \end{figure}

\begin{table*}\centering

\setlength{\tabcolsep}{4pt}
\caption{\label{tab:resultsh32} \small Derived photometric parameters from each light curve  using \gemc. Incomplete transits (denoted by $^*$) were fitted keeping the sum of the fractional radii fixed to a value of $0.1890$ in keeping with the results from \citep{Gibson2013}.  }
\begin{tabular}{lcccccc} \hline
Date & Label & Filter & Radius & Sum of            & Orbital Inclination & Transit epoch  \\
         &          &           & ratio    & fractional radii & (degrees)              & (BJD/TDB)       \\
\hline
2011/08/24 & R1 & R              & 0.1479 $\pm$ 0.0031  & 0.1918 $\pm$ 0.0070 & 89.25 $\pm$ 1.37 & 2455798.60255 $\pm$ 0.00051  \\
2011/08/24 & U1 &\textit{u} & 0.1505 $\pm$ 0.0019  & 0.1931 $\pm$ 0.0027 & 89.19 $\pm$ 0.88 & 2455798.60246 $\pm$ 0.00024  \\
2011/08/24 & B1 &\textit{b} & 0.1537 $\pm$ 0.0015  & 0.1903 $\pm$ 0.0024 & 88.81 $\pm$ 0.83 & 2455798.60239 $\pm$ 0.00020  \\
2011/08/24 & Y1 &\textit{y} & 0.1510 $\pm$ 0.0013  & 0.1904 $\pm$ 0.0025 & 88.32 $\pm$ 0.86 & 2455798.60223 $\pm$ 0.00020  \\
2011/10/04 & R2 & R              & 0.1512 $\pm$ 0.0013  & 0.1906 $\pm$ 0.0023 & 88.69 $\pm$ 0.85 & 2455839.45261 $\pm$ 0.00017  \\
2014/01/11 & V1 & V              & 0.1502 $\pm$ 0.0014  & 0.1890$^*$                & 89.34 $\pm$ 0.56  & 2456669.35548 $\pm$ 0.00037  \\
2014/08/31 & I1 & I              & 0.1553 $\pm$ 0.0017  & 0.2021 $\pm$ 0.0034  & 86.55 $\pm$ 1.10  & 2456901.55634 $\pm$ 0.00016  \\
2014/09/05 & R3 & R              & 0.1529 $\pm$ 0.0014  & 0.1877 $\pm$ 0.0019 & 89.08 $\pm$ 0.88 & 2456905.85649 $\pm$ 0.00022  \\
2014/10/24 & V2 & V              & 0.1578 $\pm$ 0.0012  & 0.1890$^*$                & 89.57 $\pm$ 0.63 & 2456955.30654 $\pm$ 0.00043  \\
2014/12/21 & V3 & V              & 0.1515 $\pm$ 0.0008  & 0.1901 $\pm$ 0.0004 & 88.94 $\pm$ 0.43 & 2457013.35687 $\pm$ 0.00008  \\
2015/08/25 & I2 & I              & 0.1512 $\pm$ 0.0007  & 0.1892 $\pm$ 0.0009  & 88.60 $\pm$ 0.50 & 2457260.60777 $\pm$ 0.00010  \\
\hline
\end{tabular}

\end{table*}

\begin{figure*} \includegraphics[width=\textwidth,angle=0]{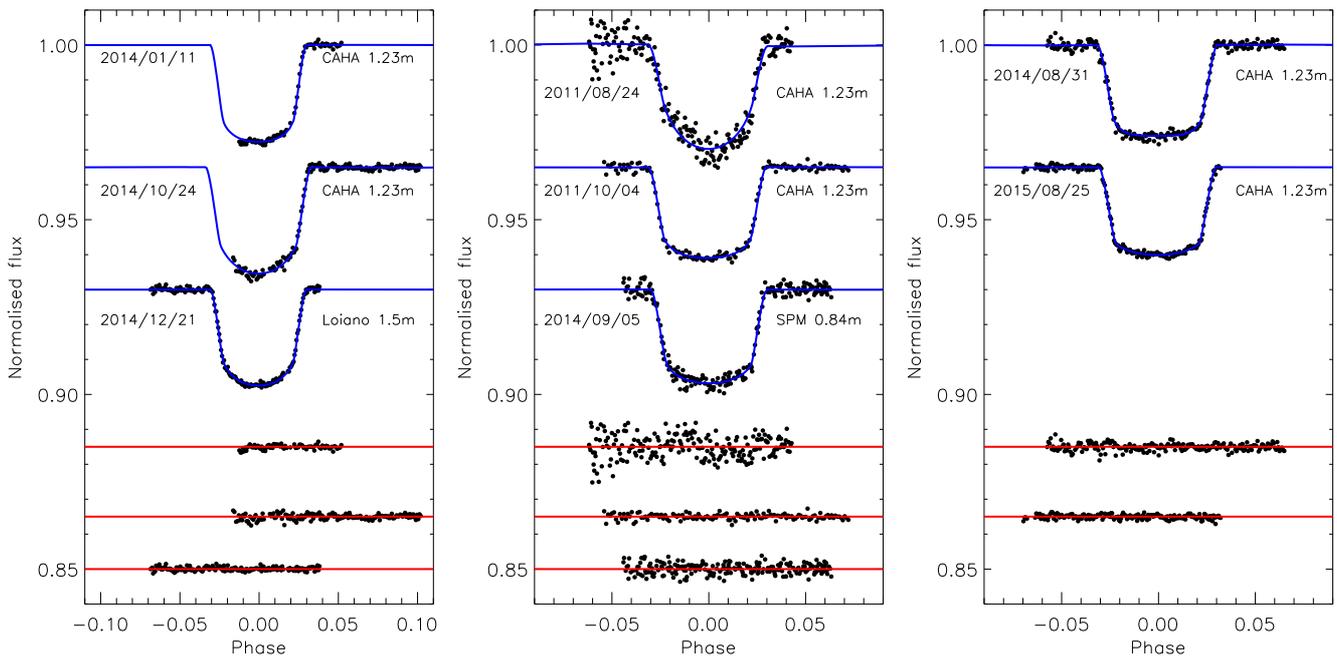}
\caption{\label{fig:all_fit_data}Transit light curves, best-fitting models and the residuals of HAT-P-32 for the eight transit light curves observed using the CAHA 1.23\,m, SPM 0.84\,cm and Cassini 1.5\,m telescopes.  Left: The three transits observed using a Johnson V filter. Middle: Three transits observed using the Cousins R and Bessell R filters. Right: Two transits observed using a Cousins I filter. The dates of the observations are on the left-hand side of each transit, and the telescope used is on the right-hand side of each transit.} \end{figure*}

All 11 transits of HAT-P-32 were modelled using \prism\ and \gemc\ (see Table\,\ref{tab:resultsh32} \& Figs.\,\ref{fig:busca-lc} \&\,\ref{fig:all_fit_data}). This was accomplished by selecting a large parameter search space to allow the global best fit solution to be found. As discussed in \citet{Jeremy2012,Jeremy2015}, the ability of \gemc\ to find the global minimum in a short amount of computing time meant that it was possible to search a large area of the parameter space to avoid the possibility of missing the best solution. Both the third light ratio and the detrending polynomial coefficient were fitted during the modelling stage. Due to a first order polynomial being used to detrend the data, only a first order polynomial was used to model the data. From a previous study of HAT-P-32 it was confirmed that the planet followed a circular orbit \citep{Zhao2014}, therefore, the orbital eccentricity ($e$) and the argument of periastron ($\omega$) were set to zero and not fitted.

For the two incomplete transits V1 and V2 the sum of the fractional radii was fixed to the value found by \citet{Gibson2013}: $0.1890$, this was done to maintain homogeneity with the results from the planetary radius variations (see Section\,\ref{Sec:rp_var}), while, also agreeing within the 1-$\sigma$ uncertainties with the remaining data sets.


\subsection{Third light ratios}
\label{sec:thirdlight}
Due to the blended M-dwarf companion (HAT-P-32\,B) within the HAT-P-32 defocussed PSF, the light ratio in the passbands in which the transits were observed needed to be found before the transits could be modelled. We used the light ratios determined by \citet{Zhao2014} in the $r'$ and $K_s$ passbands. These passbands were selected as they give the largest wavelength range from all the possible measured light ratios in \citet{Zhao2014}, thus improving the extrapolation to the passbands needed in this work. For the analysis we took the Str\"{o}mgren filter profiles from the Calar Alto observatory\footnote{See \href{https://www.caha.es/guijarro/BUSCA/caracter.html}{\begin{tt}https://www.caha.es/guijarro/BUSCA/caracter.html\end{tt}}}. The profiles from \citet{Bessell2012} were used for the V, R and I passbands as these are by design an approximation and improvement of the Johnson V, Cousins R and I filter profiles. From this analysis we determined the third light ratios (see Table\,\ref{tab:thirdlight}) needed for the different passbands used to observe the transits in this work. All the light ratios are below the 1\% flux contamination level. We then used the respective passband light ratios in \prism\ and \gemc\ to model (see Section\,\ref{sec:prism-gemc}) the 11 transits presented in this work.

\begin{table} \centering
\caption{\label{tab:thirdlight} Extrapolated third light ratios for the passbands used in the modelling of the HAT-P-32 transit light curves.}
\setlength{\tabcolsep}{8pt}
\begin{tabular}{lc} \hline
Passband                                      & Third light ratio                                              \\
\hline
Str\"{o}mgren \textit{u}  &   $0.00036\pm0.00014$        \\
Str\"{o}mgren \textit{b}  &  $0.00114\pm0.00058$        \\
Str\"{o}mgren \textit{y}  & $0.00213\pm0.00088$        \\
Johnson V                         &  $0.00212\pm0.00084$       \\
Cousins R                         & $0.00354\pm0.00132$ \\
Bessell R                            & $0.00354\pm0.00132$ \\
Cousins I                         & $0.00714\pm0.00154$ \\
\hline \end{tabular}
\end{table}


\subsection{Photometric results}
\label{Sec:photoresults}

The photometric parameters for the HAT-P-32 system are given in Table\,\ref{tab:resultsh32}. The weighted means of the system parameters and with their 1-$\sigma$ uncertainties together with their comparisons to published values are given in Table\,\ref{tab:resultsh32final}. The combined photometric results shows excellent agreement with previous published results. Fig.\,\ref{fig:busca-lc} \& Fig.\,\ref{fig:all_fit_data} compares the light curves to the best-fitting models, including the residuals.

\begin{table}
\caption{\label{tab:resultsh32final} Combined photometric parameters for HAT-P-32, compared to the values found by \citet{Hartman2011}\,(H11), \citet{Sada2012}\,(S12), \citet{Gibson2013}\,(G13), \citet{Seeliger2014}\,(S14), \citet{Mallonn2016a}\, blue (M16\,B), red (M16\,R) and \citet{Nortmann2016}\,(N16).
The photometric parameters are the weighted means from the data sets, which, have measured uncertainties.}
\setlength{\tabcolsep}{4pt}
\begin{tabular}{lccc} \hline
                              & $r_p / r_*$                                       & $r_* + r_p$                              & $i$                                              \\
\hline
This                    &                                               &                                                 &                                                      \\
Work                       &  {\bf0.1515 $\pm$ 0.0004} & {\bf 0.1902 $\pm$ 0.0003}     & {\bf 88.98 $\pm$ 0.21 }             \\
\hline
H11                       &   0.1508 $\pm$ 0.0004        & 0.1902 $\pm$ 0.0005$^*$                             & 88.9 $\pm$ 0.4                           \\
S12                        &  0.1531 $\pm$ 0.0012        &   0.1928 $\pm$ 0.0029$^*$                           & 88.16$^{+1.03}_{-1.17}$         \\
G13                        & 0.1515 $\pm$ 0.0012         &  0.1890 $\pm$ 0.0015$^*$                            & 89.12$^{+0.61}_{-0.68}$         \\
S14                        &  0.1510 $\pm$ 0.0004        &  0.1901 $\pm$ 0.0005$^*$                            & 88.92 $\pm$ 0.10                        \\
M16\,B            &  0.1515 $\pm$ 0.0012        &  0.1904 $\pm$ 0.0030$^*$                            & 88.61 $\pm$ 0.84                        \\
M16\,R             &  0.1505 $\pm$ 0.0005        &  0.1903 $\pm$ 0.0018$^*$                            & 88.56 $\pm$ 0.57                        \\
N16                       &  0.1516$^{+0.0009}_{-0.0005}$        &  0.1881$^{+0.0018}_{-0.0007}$ $^*$          & 89.33$^{+0.58}_{-0.80}$                       \\
\hline \end{tabular}
\begin{footnotesize}\hspace*{0cm} $^*$The sum of the fractional radii from the literature was calculated using the respective values for $R_p / R_*$ and $a / R_*$. \end{footnotesize}
\end{table}

\begin{table} \begin{center}
\caption{\label{tab:minima} Times of minimum light of HAT-P-32
and their residuals versus the ephemeris derived in this work.
\newline {\bf References:}
(1) \citet{Hartman2011};
(2)  \citet{Sada2012};
(3) \citet{Gibson2013};
(4)  \citet{Seeliger2014};
(5) \citet{Mallonn2016a};
(6)  \citet{Nortmann2016};
(7) This Work
}
\vspace{0.3cm}
\begin{tabular}{l@{\,$\pm$\,}l r r c} \hline
\multicolumn{2}{l}{Time of minimum}   & Cycle  & Residual & Reference \\
\multicolumn{2}{l}{(BJD/TDB $-$ 2400000)} & no.    & (BJD)    &           \\
\hline
54420.44712 & 0.00009 &     0.0 & -0.00007 &  1 \\   
55798.60255 & 0.00051 &   641.0 &  0.00023 &  7 \\   
55798.60246 & 0.00024 &   641.0 &  0.00014 &  7 \\   
55798.60239 & 0.00020 &   641.0 &  0.00007 &  7 \\   
55798.60223 & 0.00020 &   641.0 & -0.00009 & 7  \\   
55839.45261 & 0.00017 &   660.0 &  0.00014 &  7 \\   
55843.75341 & 0.00019 &   662.0 &  0.00092 &  2 \\   
55845.90287 & 0.00024 &   663.0 &  0.00038 &   2\\   
55845.90314 & 0.00040 &   663.0 &  0.00065 &   2\\   
55867.40301 & 0.00073 &   673.0 &  0.00044 &  4 \\   
55880.30267 & 0.00033 &   679.0 &  0.00005 &   4\\   
55895.35297 & 0.00016 &   686.0 &  0.00029 &   4\\   
55895.35249 & 0.00080 &   686.0 & -0.00019 &   4\\   
55897.50328 & 0.00033 &   687.0 &  0.00059 &   4\\   
55910.40274 & 0.00043 &   693.0 &  0.00001 &  4 \\   
55923.30295 & 0.00031 &   699.0 &  0.00017 &   4\\   
55942.65287 & 0.00064 &   708.0 &  0.00002 &   4\\   
56155.50385 & 0.00026 &   807.0 &  0.00020 &   4\\   
56157.65470 & 0.00072 &   808.0 &  0.00105 &   4\\   
56177.00392 & 0.00025 &   817.0 &  0.00020 &  3 \\   
56183.45364 & 0.00085 &   820.0 & -0.00011 &   4\\   
56183.45361 & 0.00049 &   820.0 & -0.00014 &   4\\   
56185.60375 & 0.00033 &   821.0 & -0.00001 &   4\\   
56185.60379 & 0.00011 &   821.0 &  0.00003 &   6\\   
56211.40361 & 0.00056 &   833.0 & -0.00024 &   4\\   
56220.00440 & 0.00019 &   837.0 &  0.00051 &  3 \\   
56245.80345 & 0.00007 &   849.0 & -0.00053 &  5 \\   
56254.40404 & 0.00022 &   853.0 &  0.00003 &   4\\   
56542.50538 & 0.00032 &   987.0 &  0.00029 &   4\\   
56542.50530 & 0.00018 &   987.0 &  0.00021 &   4\\   
56542.50522 & 0.00052 &   987.0 &  0.00013 &   4\\   
56572.60532 & 0.00018 &  1001.0 &  0.00012 &   4\\   
56598.40539 & 0.00017 &  1013.0 &  0.00009 &   4\\   
56600.55546 & 0.00017 &  1014.0 &  0.00016 &   4\\   
56628.50585 & 0.00031 &  1027.0 &  0.00044 &   4\\   
56656.45533 & 0.00045 &  1040.0 & -0.00018 &   4\\   
56669.35548 & 0.00037 &  1046.0 & -0.00008 &   7\\   
56901.55634 & 0.00016 &  1154.0 & -0.00008 &  7 \\   
56905.85649 & 0.00022 &  1156.0 &  0.00005 &  7 \\   
56955.30654 & 0.00043 &  1179.0 & -0.00008 &  7 \\   
57013.35687 & 0.00008 &  1206.0 &  0.00003 &  7 \\   
57260.60777 & 0.00010 &  1321.0 &  0.00001 &  7 \\   
\hline \end{tabular} \end{center} \end{table}

\begin{figure*} \includegraphics[width=\textwidth,angle=0]{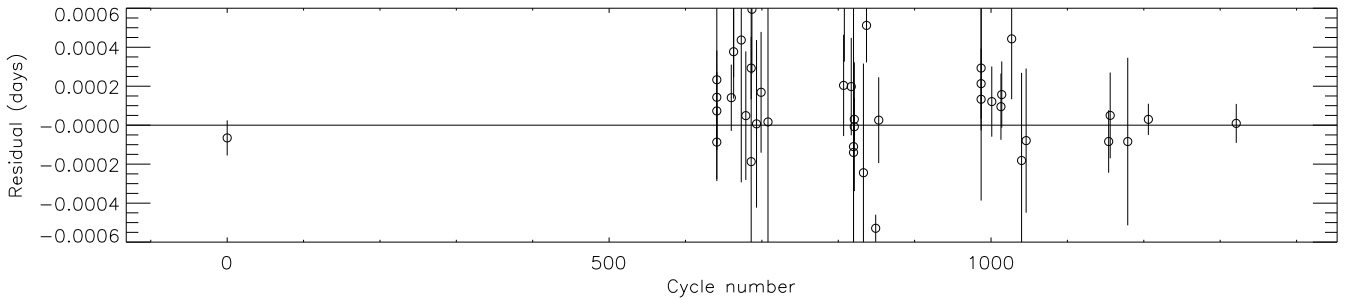}
\caption{\label{fig:oc} Residuals of the available times of mid-transit
versus the orbital ephemeris found for HAT-P-32. } \end{figure*}

The available times of mid-transit for HAT-P-32 were collected (see Table\,\ref{tab:minima}) from the literature \citep{Hartman2011,Sada2012,Gibson2013,Seeliger2014,Mallonn2016a,Nortmann2016}. The value used from \citet{Mallonn2016a} was calculated as the weighted mean between the independently fitted blue and red values. All timings were converted to the BJD/TDB timescale and used to obtain an improved orbital ephemeris: $$ T_0 = {\rm BJD/TDB} \,\, 2\,454\,420.447187 (96) \, + \,2.15000800 (10) \times E $$ where $E$ represents the cycle count with respect to the reference epoch and the bracketed quantities represent the uncertainty in the final two digits of the preceding number. Fig.\,\ref{fig:oc} and Table\,\ref{tab:minima} show the residuals of these times against the ephemeris. The overall fit of the times of mid-transit are in agreement with a linear ephemeris by 1.6-$\sigma$, which, indicate that the results show no evidence for transit timing variations. When the two major outliers, at 7.6-$\sigma$ \citep[56245.80345:][]{Mallonn2016a} and 4.9-$\sigma$ \citep[55843.75341:][]{Sada2012}, are removed from the analysis the overall fit improves to 0.9-$\sigma$.

The times of mid-transit from \citet{Seeliger2014} were taken from the 20 `good' transits presented in their work. However, the transit they  obtained on 2013/01/04 was not used in this analysis due to the reported mid-transit time not agreeing with the reported date.

\subsection{Physical properties of the HAT-P-32 planetary system}
\label{Sec:phyiscalresults}

\begin{table} \centering
\caption{\label{tab:modelh32} Physical properties of the HAT-P-32 system. Where two errorbars are
given, the first is the statistical uncertainty and the second is the systematic uncertainty. The values found by \citet{Hartman2011} (H11) are given for comparison.}
\begin{tabular}{l r@{\,$\pm$\,}c@{\,$\pm$\,}lc}
\hline
Parameter & \mcc{This Work} & H11 \\
\hline
$M_{\rm A}$    (\Msun) & 1.182    & 0.041    & 0.026   & 1.160\,$\pm$\,0.041   \\
$R_{\rm A}$    (\Rsun) & 1.225    & 0.015    & 0.009    & 1.219\,$\pm$\,0.016   \\
$\log g_{\rm A}$ (cgs) & 4.3349   & 0.0054   & 0.0032   & 4.33\,$\pm$\,0.01    \\
$\rho_{\rm A}$ (\psun) & \mcc{$0.6435 \pm 0.0032$}   &      \\[2pt]
$M_{\rm b}$    (\Mjup) & 0.80     & 0.14     & 0.01  & 0.86\,$\pm$\,0.164      \\
$R_{\rm b}$    (\Rjup) & 1.807    & 0.022    & 0.013   & 1.789\,$\pm$\,0.025    \\
$g_{\rm b}$     (\mss) & \mcc{$6.0 \pm 1.1$}        & 6.6$^{+1.2}_{-1.4}$       \\
$\rho_{\rm b}$ (\pjup) & 0.126    & 0.023    & 0.001 & 0.143\,$\pm$\,0.030      \\[2pt]
\Teq\              (K) & \mcc{$1801 \pm   18$}     & 1786\,$\pm$\,26     \\
\safronov\             & 0.0256   & 0.0046   & 0.0002   & 0.028\,$\pm$\,0.005   \\
$a$               (AU) & 0.03448  & 0.00041  & 0.00025 & 0.0343\,$\pm$\,0.0004    \\
Age              (Gyr) & \ermcc{2.2}{0.7}{0.7}{0.5}{0.3} & 2.7\,$\pm$\,0.8  \\
\hline \end{tabular}  \end{table}

We used the same approach\footnote{For a detailed discussion on the methodology used in the analysis see \citet{Me09mn}.} as described by \citet{Jeremy2015}, in that we used the photometric properties of HAT-P-32 to determine the physical characteristics. The analysis used a set of parameters which were obtained from the analysed light curves and previously published spectra, plus tabulated predictions of theoretical models. We adopted the values of $i$, $r_{\rm p}/r_\star$ and $r_\star + r_{\rm p}$ from Table\,\ref{tab:resultsh32final}, while, the orbital velocity amplitude $K_{\star} = 110\pm16$\ms, the stellar effective temperature $\Teff = 6269\pm64$\,K and metal abundance $\FeH = -0.04 \pm 0.08$ from \citet{Zhao2014}.

The standard formulae and the physical constants listed by \citet{Me11mn} were used in conjunction with a starting value of $K_{\rm p}$, to calculate the physical properties of the system. The stellar expected \Teff\ and radius was determined through interpolating the mass and \FeH\ of the star within a set of tabulated predictions from theoretical stellar models. At each iteration $K_{\rm p}$ was refined until the best agreement was found between the expected and observed \Teff, and the expected $\frac{R_{\rm \star}}{a}$ and measured $r_{\rm \star}$. This was performed from the zero-age to the terminal-age main sequence, in steps of 0.01\,Gyr. This approach then yielded the estimates of the system physical parameters and the evolutionary age of the star.

Due to the differing agreements and systematic errors between various theoretical stellar models, this methodology was repeated separately using five different sets of stellar theoretical models \citep[see][]{Me10mn}, and the Gaussian distribution of the parameter output values was used to determine the systematic error. A perturbation algorithm was then used to propagate the statistical errors \citep[see][]{Me10mn}.

The final results of this process are in agreement with themselves and are in excellent agreement with published results for HAT-P-32 (see Table\,\ref{tab:modelh32}). The final physical properties are given in Table\,\ref{tab:modelh32} and contains the individual statistical and systematic errorbars for the parameters which have a dependency on the theoretical models. The largest of the five statistical errorbars from the five theoretical stellar models, is used for the final statistical errorbar, for each parameter. The same is true for the systematic errorbar which is calculated from the standard deviation (1-$\sigma$) of the parameter values.


\section{Variation of planetary radius with wavelength}
\label{Sec:rp_var}

\begin{table*} \centering

\setlength{\tabcolsep}{4pt}
\caption{\label{tab:rp-results} Values of $r_p$ and $R_p$ for each light curve. The uncertainties exclude all common uncertainties in $r_p$ and $R_p$, and so, should only be used to compare different values of $r_p\left(\lambda\right)$ and $R_p\left(\lambda\right)$. The final column gives the uncertainty in $R_p$ in units of the atmospheric scale height, $H$.  }
\begin{tabular}{lccccccc} \hline
Telescope /        & Label & Passband                        & $\lambda_{cen}$ & FWHM    & $r_p$        & $R_p$            & $\sigma\left(H\right)$  \\
Instrument         &           &                             & (nm)                     & (nm)        &                   & (\Rjup)          &                                      \\
\hline
BUSCA               & U1 & Str\"{o}mgren \textit{u} & 350                      & 30            & 0.02513 $\pm$ 0.00008 & 1.850 $\pm$ 0.006 & 0.39   \\
BUSCA               & B1 & Str\"{o}mgren \textit{b} & 467                      & 18            & 0.02497 $\pm$ 0.00007 & 1.838 $\pm$ 0.005 & 0.35  \\
BUSCA               & Y1 & Str\"{o}mgren \textit{y} & 547                      & 23            & 0.02486 $\pm$ 0.00007 & 1.830 $\pm$ 0.005 & 0.36   \\
CAHA\,1.23\,m  & V1 & Johnson V                       & 544.8                  & 84             & 0.02496 $\pm$ 0.00007 & 1.837 $\pm$ 0.005 & 0.33    \\
CAHA\,1.23\,m  & V2 & Johnson V                       & 544.8                  & 84              & 0.02592 $\pm$ 0.00009 & 1.908 $\pm$ 0.007 & 0.43    \\
Cassini 1.5\,m    & V3 & Johnson V                       & 544.8                  & 84              & 0.02489 $\pm$ 0.00005 & 1.832 $\pm$ 0.004 & 0.23   \\
CAHA\,1.23\,m  & R1 & Cousins R                        & 640.7                  & 158          & 0.02423 $\pm$ 0.00041 & 1.784 $\pm$ 0.030 & 1.96 \\
CAHA\,1.23\,m  & R2 & Cousins R                        & 640.7                  &158           & 0.02468 $\pm$ 0.00007 & 1.816 $\pm$ 0.005 & 0.33   \\
SPM\,0.84\,m    & R3 & Bessell R                          & 630                      & 118           & 0.02506 $\pm$ 0.00009 & 1.845 $\pm$ 0.006 & 0.42   \\
CAHA\,1.23\,m  & I2 & Cousins I                         & 798                     & 154           & 0.02491 $\pm$ 0.00006  & 1.834 $\pm$ 0.004 & 0.27    \\
\hline
\end{tabular}

\end{table*}

The 11 datasets of the HAT-P-32 planetary system presented in this work were obtained using seven different passbands. One dataset was observed simultaneously in three-passbands (Str\"{o}mgren \textit{u}, \textit{b} \& \textit{y} from BUSCA), and the other eight were observed using Johnson V, Bessell R, Cousins R and I passbands (from the CAHA 1.23\,m, SPM 0.84\,m and Cassini 1.5\,m telescopes). Due to this, it is only natural to search for possible variations in the planetary radius in these passbands. For this we follow the same procedure of \citet{Sou2015}, in that we refit the light curves with all the parameters fixed, except for $k$, $T_0$ and the detrending polynomial coefficients. We keep $T_0$ fixed for the two incomplete transits (V1 and V2). As mentioned in Section\,\ref{sec:tele} the I1 transit was not used in this analysis, so only ten datasets were used.

The fractional planetary radius, $r_p$  is represented in our modelling of the light curves by the parameter $k$, which is directly linked to the primary observable: the transit depth. The parameter which is directly comparable to theoretical predictions is the absolute planetary radius ($R_p$). In \prism\ the fractional radii are used\footnote{The fractional radii share a correlation with the other photometric parameters \citep[see][]{Me08mn}}, so a transformation using the semimajor axis, $a$ is required to find $R_p$ from $r_p$: $R_p = a \cdot r_p $. However, $a$ (and its associated uncertainty) is an absolute property of the system and therefore will be the same, irrespective of which passband is used to observe a transit. Refitting the light curves by using a fixed $a$ allows to find a set of \Rjup\ values and uncertainties which are directly comparable to each other (see Table\,\ref{tab:rp-results}).

Our ten planetary radius measurements cover the optical wavelength range from 350\,nm to 798\,nm. In order to increase the scope of our analysis we include the results from \citet{Gibson2013} (520--930\,nm). To obtain a direct comparison between the two sets of results, we fixed the fractional stellar radius and the orbital inclination to the values found by \citet{Gibson2013}, when we refitted the light curves. This was made possible due to the fact that our results for these two parameters are in agreement to the values from \citet{Gibson2013} (see Table\,\ref{tab:resultsh32final}).

Figs.\,\ref{fig:rp_lambda1}\,and\,\ref{fig:rp_lambda2} show the transformed $R_p$ values as a function of the central wavelength of the passband from the analysis. The FWHM of each passband is shown as a horizontal line for reference. The fitted parameter $r_p$ and the passband characteristics are given in Table\,\ref{tab:rp-results} together with the uncertainties in $R_p$ given in units of pressure scale height, $H$. We calculated $H$ using the planetary equilibrium temperature, $1801\pm18$\,K (see Table\,\ref{tab:modelh32}) and found an agreement with the approximation ($H\approx 1100$\,km) given by \citet{Gibson2013}, with $H=1070\pm170$\,km. The relative uncertainties for 90\,\% of our measured radii of HAT-P-32\,A\,b are smaller than one atmospheric pressure scale height. This indicates that our measurements are sensitive to radius variations at the 1\,$H$ level. Our data therefore, are in principle, sensitive to the properties of the atmosphere of HAT-P-32\,A\,b.

By examining Table\,\ref{tab:rp-results} it can be seen that the refitted planetary radius from dataset V2 is larger than expected (considering the values from the remaining Johnson V passbands). As this is one of the partial transits the anomalous result can be explained as an artefact from the data reduction stage. We therefore did not use the $R_p$ from this transit in the comparison to the theoretical model spectra. The transit observed with the Cousins R filter on the CAHA 1.23\,m telescope, R1 appears to be shallower than expected. This smaller radius can be accounted for by the poor quality of the light curve, due to the contribution of systematics and the small amount of defocus used. We therefore used the weighted mean of $R_p$, from the two observed transits in the Cousins R filter (R1 \& R2) for our comparisons to the theoretical model spectra.


\subsection{Theoretical transmission model spectra}
\label{Sec:rp_var_model}

\begin{figure*} \includegraphics[width=\textwidth,angle=0]{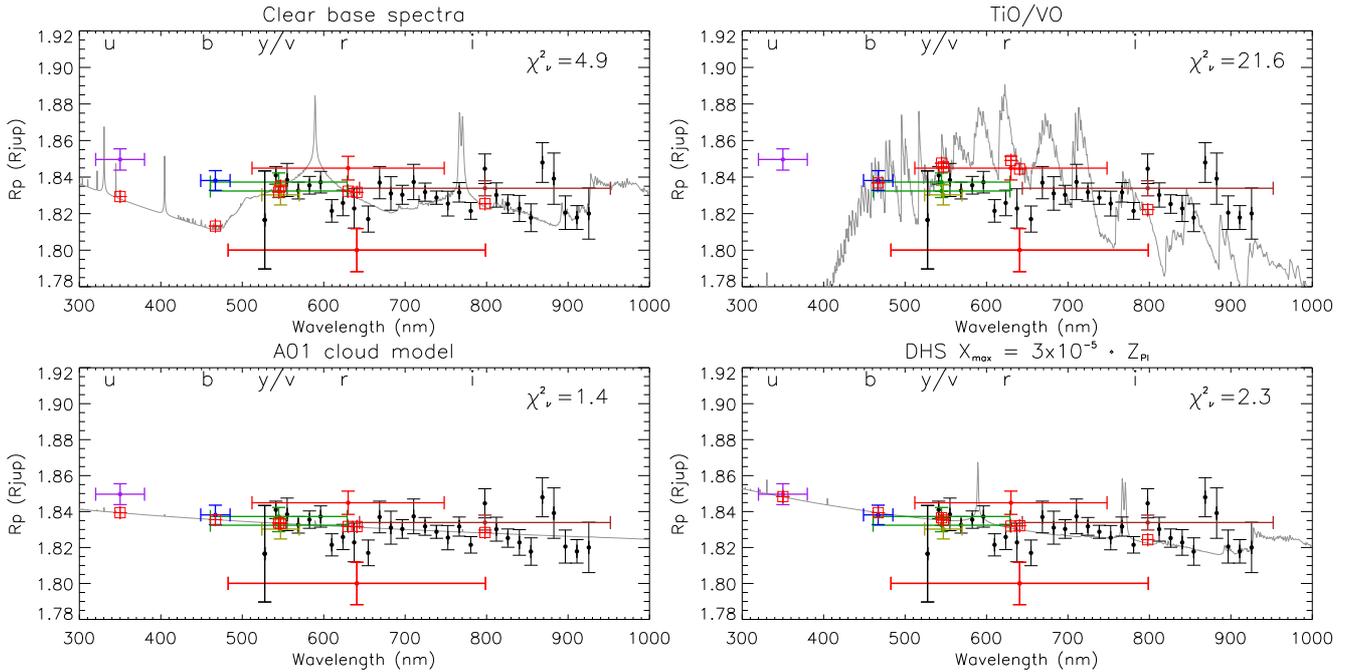}
\caption{\label{fig:rp_lambda1}Measurements of the planetary radius compared to predicted theoretical model atmospheres from the {\tt petitCODE} \citet{Molliere2015, Molliere2017}. The data points show the measured $R_p$ from each light curve, where the vertical error bars represent the relative uncertainty in $R_p$, and the horizontal error bars represent the FWHM of the corresponding passband. The models are represented separately in each plot, with the plot title giving the model spectra. Eight data points from this work (see Table\,\ref{tab:rp-results}) are represented based on their passband colour and the \citet{Gibson2013} data points are black. The red open squares are the passband averages of the transmission spectra models, and are shown at the central wavelengths of the relevant passbands. The general band names (i.e., r-band) are displayed at the top of each plot along with the best fitting $\chi^2_{\nu}$ for each model spectrum. } \end{figure*}

We initially compared our planetary radius measurements to 15 theoretical transmission spectra which, were generated\footnote{Three additional transmission spectra were generated as variations of a bimodal cloud particle distribution by altering different atmospheric model parameters.} by the model atmosphere code of \citet{Molliere2015, Molliere2017}, seven of which are shown in Figs.\,\ref{fig:rp_lambda1}\,and\,\ref{fig:rp_lambda2}. {\tt petitCODE} \citep{Molliere2015, Molliere2017} is a model which calculates exoplanet atmospheric structures in radiative--convective equilibrium, including absorption and scattering processes, and the self--consistent treatment of clouds. As an output the code returns the planet's emission and transmission spectra. For the model calculations presented here, a two--pronged approach for generating cloudy spectra was followed: {\bf(i)} using the planet--star system parameters (host star temperature and radius, planet's semi--major axis, radius and mass), and assuming a fiducial atmospheric enrichment of [Fe/H] = 0.55 we calculated self-consistent clear and cloudy structures and spectra for HAT-P-32\,A\,b. For these calculations the cloud model parameter choice as defined in \citet{Molliere2017}, Table\,2, was used, while the atmospheric enrichment was chosen following the method described in Section 4.1 of \citet{Molliere2017}. {\bf(ii)} In addition to these spectra with a self--consistent cloud treatment we also considered the standard approach \citep[see, e.g.,][]{Sing2016} to take our fiducial cloud-free atmospheric structure of this planet, and adding a grey cloud deck and/or a Rayleigh scattering haze, the latter of which was included by scaling the H$_2$/He Rayleigh cross-sections by a given factor.

Cloud modelling following approach {\bf(i):} in the self--consistent cloud calculations, the particle opacities for the clouds are determined from either Mie theory or the distribution of hollow spheres (DHS) approach \citep{Min2005}. Mie theory uses the classical assumption of  spherically homogeneous grains. DHS uses a distribution of hollow spheres to determine the optical properties of irregularly shaped dust aggregates. The model assumes zero interaction between the different chemical species of clouds. The different species considered are MgAl$_2$O$_4$, Mg$_2$SiO$_4$, Fe, KCl and Na$_2$S \citep{Molliere2017}. 

The 15 theoretical transmission spectra generated and used in this work span a range of different atmospheric model parameters: metal enrichment, C/O number ratio, TiO/VO opacity, cloud particle settling parameter, cloud mass fractions, cloud deck pressures and Rayleigh haze scaling factors\footnote{As written above, in the calculations in which the clouds are not included in a self-consistent fashion the use of a Rayleigh scattering haze does not stem from H$_2$/He, however, it is how the haze is implemented: small particle clouds (i.e., hazes) with particle sizes smaller than the observation wavelength, lead to a Rayleigh scattering signal. But as the cloud species are unknown, the H$_2$/He cross--sections are scaled, to mimic the haze.}. The first theoretical transmission spectrum which was generated, represents a clear cloudless model using a scaled solar metal enrichment level $\left(\FeH = 0.55\right)$ combined with a solar C/O number ratio. TiO/VO opacity was not added. This spectrum can be considered as the `base' spectrum in this work. Five more base transmission spectra were generated with the following variations: an order of magnitude increase and decrease in the metal enrichment (e.g., $\FeH = -0.45$ and $\FeH = 1.55$), and a doubling and halving of the C/O number ratio. TiO/VO opacity was added to the fifth base transmission spectrum.

Cloud opacity was added to a further six generated base transmission spectra. The cloud opacity was treated using the Self Consistent Coupling (SCC) method as described by \citet{Molliere2017}. One of the cloud transmission spectra used the \citet{Ackerman2001} (A01) cloud model to allow the coupling between the effects of clouds with the atmospheric temperature iteration. It should be noted that for this work the implementation of the \citet{Ackerman2001} model differs from the original, by accounting for the vertical mixing induced by insolation and setting the radiative layer mixing length equal to the atmospheric pressure scale height \citep[see][]{Molliere2017}. For this transmission spectrum DHS was used to describe the cloud particles with the cloud particle settling parameter set to: $f_{sed} = 1$, which is the ratio between the mass averaged grain settling velocity and the atmospheric mixing velocity \citep{Molliere2017}.  It was found by \citet{Molliere2017} that it is only possible to replicate a steep Rayleigh slope in the optical, if small cloud particles ($\approx$0.06 to 0.12\,$\mu$m) are placed into the atmosphere at high layers. Therefore the other five base cloudy transmission spectra using the SCC method used a parametrised cloud model, corresponding to vertically homogeneous clouds, however, not larger than a given maximum value, which is a simple way of treating settling, and used a mono--disperse particle size of 0.08\,$\mu$m. The first spectrum used Mie theory (homogeneous spherical grains) to describe the cloud particles and used a parametrised cloud model with a maximum cloud mass fraction within the atmosphere set to: $X_{max} = 3\times10^{-4} \cdot Z_{Pl}$ (where $Z_{Pl}$ is the atmospheric metal mass fraction). Three spectra were generated using DHS to describe the cloud particles but each had a different maximum cloud mass fraction: $X_{max} = 10^{-2} \cdot Z_{Pl}$, $3\times10^{-4} \cdot Z_{Pl}$ and  $3\times10^{-5} \cdot Z_{Pl}$. The final base cloudy transmission spectrum generated using the SCC method had a maximum cloud mass fraction of $X_{max} = 3\times10^{-4} \cdot Z_{Pl}$, however, Fe opacity was added to the clouds. This has the effect of dampening any Rayleigh scattering in the optical due to the strong absorbing nature of Fe in the optical \citep{Molliere2017}. 

Cloud modelling following approach {\bf(ii):} three theoretical transmission spectra were generated by using the cloudless self-consistent atmospheric structures, obtained as described above, and then adding cloud opacity only for the spectral calculations. The properties of the added cloud opacity are determined from the cloud pressure deck and a Rayleigh haze scaling factor. Each of the three transmission spectra had a metal enrichment level: $\FeH = 0.55$ combined with a solar C/O number ratio. One of the spectra had a cloud pressure deck set at 0.001\,bar. For the second transmission spectrum a Rayleigh haze scaling factor of 100 was used with an omitted cloud pressure deck. The final transmission spectrum had both a 0.001\,bar cloud pressure deck and an added Rayleigh haze scaling factor of 100.


\subsection{Fitted theoretical transmission spectra results}
\label{Ray and cloud}

\begin{figure*} \includegraphics[width=\textwidth,angle=0]{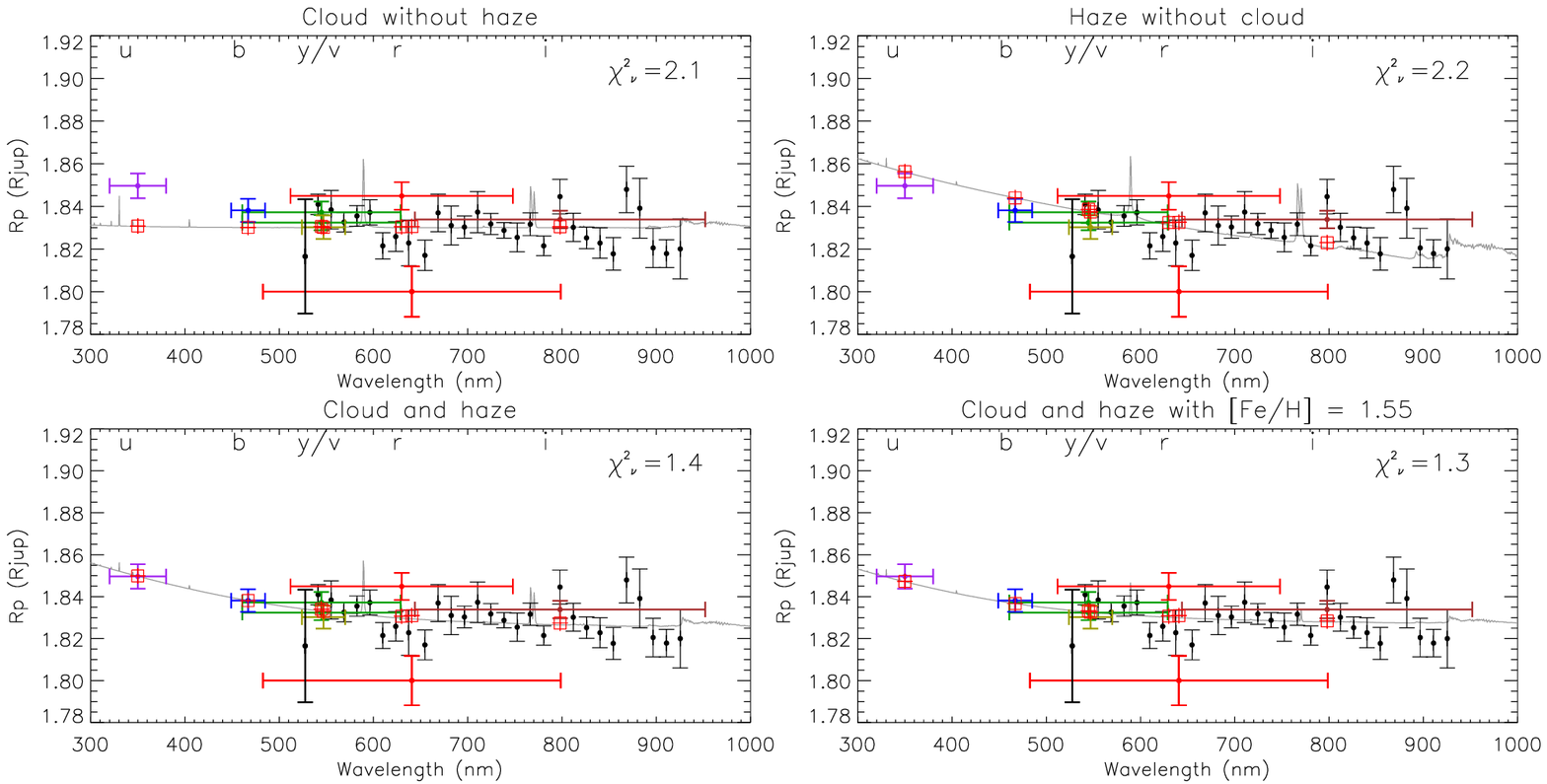}
\caption{\label{fig:rp_lambda2}Measurements of the planetary radius compared to predicted theoretical model atmospheres from the {\tt petitCODE} \citet{Molliere2015, Molliere2017}. The data points show the measured $R_p$ from each light curve where, the vertical error bars represent the relative uncertainty in $R_p$ while, the horizontal error bars represent the FWHM of the corresponding passband. The models are represented separately in each plot, with the plot title giving the model spectra. Eight data points from this work (see Table\,\ref{tab:rp-results}) are represented based on their passband colour and the \citet{Gibson2013} data points are black. The red open squares are the passband averages of the transmission spectra models, and are shown at the central wavelengths of the relevant passbands. The general band names (i.e., r-band) are displayed at the top of each plot along with the best fitting $\chi^2_{\nu}$ for each model spectra.} \end{figure*}

We fitted the radial offset of each model spectrum to the planetary radius measurements via a MCMC algorithm and determined the reduced chi squared, $\chi^2_\nu$ between all the theoretical transmission spectra and the planetary radius measurements, whilst taking into account the FWHM of each passband.

\begin{equation}
\chi^2_{\nu} = \frac{\chi^2}{\left(n-\theta\right)}
\end{equation}

\noindent where $\chi^2$ is the chi squared value, $n$ is the total number of data points, $\theta$ is the number of fitted model parameters, and so, $\left(n-\theta\right)$ is the number of degrees of freedom\footnote{The number of degrees of freedom ($dof$) used in this work was $dof = 36$.}.

\begin{table} \begin{center}
\caption{\label{tab:rp_lambda} Best fit statistics for the theoretical transmission spectra and the planetary radius measurements. The theoretical transmission spectra are split into three distinct groups; Cloudless clear spectra, cloudy spectra using the SCC (Self Consistent Coupling) method and cloudy spectra by adding cloud opacity. The two best fitting spectra are highlighted in bold.
}
\setlength{\tabcolsep}{6pt}
\begin{tabular}{lcc} \hline
Model spectra                                     &  Best Fit    & BUSCA agreement                                              \\
					& $\left(\chi^2_\nu\right)$ & $\left(\sigma\right)$                         \\
\hline
{\bf Cloudless clear spectra} & & \\
Base					  & 4.9 & 3.0    \\
TiO/VO				  & 21.6 & 7.3    \\
$\left[\rm{Fe/H}\right] = 1.55$  & 4.5 & 3.9 \\
$\left[\rm{Fe/H}\right] = -0.45$  & 4.5 & 1.5 \\
Twice solar C/O ratio		& 5.6 & 2.8 \\
Half solar C/O ratio			& 4.9 & 3.3 \\
\hline
{\bf SCC cloud spectra} & & \\
A01 cloud model & {\bf 1.4} & {\bf 0.94} \\
Mie cloud particles$^\dagger$ & 1.6 & 0.63 \\
DHS $X_{max} = 10^{-2} \cdot Z_{Pl}$ & 1.7 & 1.2 \\
DHS $X_{max} = 3\times10^{-4} \cdot Z_{Pl}$ & 1.8 & 1.3 \\
DHS $X_{max} = 3\times10^{-5} \cdot Z_{Pl}$ & 2.3 & 0.46 \\
DHS Fe clouds$^\dagger$ & 2.1 & 1.4 \\
\hline
{\bf Added cloud opacity} & & \\
Cloud only & 2.1 & 1.6 \\
Haze only & 2.2 & 1.2 \\
Cloud and haze & {\bf 1.4} & {\bf 0.19} \\
\hline \end{tabular} \end{center}
\begin{footnotesize}\hspace*{0cm}$\dagger$\,The Mie and DHS Fe cloud models have an $X_{max} = 3\times10^{-4} \cdot Z_{Pl}$, see Table\,2 \citet{Molliere2017}. In addition, the DHS  Fe clouds model also has all the other cloud species in it, but with the addition of Fe, while the nominal DHS/Mie-$X_{max}$ models have no Fe included. \end{footnotesize}
\end{table} 

The $\chi^2_\nu$ results from fitting the 15 theoretical transmission spectra to the planetary radius measurements are presented in Table\,\ref{tab:rp_lambda}.

Fig.\,\ref{fig:rp_lambda1} shows the comparison between the best fit of four of the theoretical transmission spectra and the planetary radius measurements. The two upper panels of Fig.\,\ref{fig:rp_lambda1} show two of the clear cloudless transmission spectra: the base transmission spectrum and the transmission spectrum with added TiO/VO opacity. The two bottom panels of Fig.\,\ref{fig:rp_lambda1} show two of the cloud spectra which were generated using the SCC method: the A01 cloud model \citep{Ackerman2001} and the cloud spectrum generated using DHS to describe the cloud particles and using a maximum cloud mass fraction, $X_{max} = 3\times10^{-5} \cdot Z_{Pl}$. 

Fig.\,\ref{fig:rp_lambda2} shows the comparison between the best fit of four of the theoretical transmission spectra with added cloud opacity\footnote{Three of these spectra can be found in Table\,\ref{tab:rp_lambda}. The bottom right panel of Fig.\,\ref{fig:rp_lambda2} shows a spectrum with an added bimodal cloud opacity generated, using an alternate set of parameters (see Section\,\ref{bimodal}).}; and with that of the planetary radius measurements. The two upper panels of Fig.\,\ref{fig:rp_lambda2} show two transmission spectra: a spectrum with an added cloud pressure deck set at 0.001\,bar and the transmission spectrum with an added Rayleigh haze scaling factor of 100. The two bottom panels of Fig.\,\ref{fig:rp_lambda2} show two transmission spectra where a bimodal cloud opacity was added. 

The best fitting theoretical transmission spectrum to the planetary radius measurements (in Table\,\ref{tab:rp_lambda}) is the base spectrum with clouds from the cloud approach {\bf(i)}, generated using the A01 cloud model ($\chi^2_{\nu} = 1.4$). Equally in agreement though, at $\chi^2_{\nu} = 1.4$, is the fitted spectrum generated with approach {\bf(ii)}, i.e., a model with a metal enrichment of $\FeH = 0.55$, an added cloud opacity using a cloud pressure deck of 0.001\,bar, and a Rayleigh haze scaling factor of 100. This confirms and agrees with previous studies \citep[e.g.,][]{Gibson2013, Mallonn2016b} in detecting a grey absorbing cloud deck within the atmosphere of HAT-P-32\,A\,b. When the two best fitting spectra are examined (bottom left of Figs.\,\ref{fig:rp_lambda1} and\,\ref{fig:rp_lambda2}), it is seen that both have the same agreement over the entire wavelength range (350\,nm--798\,nm) of radius measurements. However, when the two spectra are compared to the BUSCA data (350\,nm--547\,nm) alone, it can be clearly seen that the combined cloud deck and haze spectrum gives a superior agreement at: 0.19-$\sigma$ compared with 0.94-$\sigma$ for the A01 cloud model spectrum (see Table\,\ref{tab:rp_lambda}). The detection of a Rayleigh--like scattering haze between 350\,nm--547\,nm agrees with the previous study by \citet{Mallonn2016a}. 


\subsection{Theoretical transmission spectra with added bimodal cloud opacity}
\label{bimodal}

The ensemble of different wavelength dependent radii variations used in this work were observed independently on different nights, with the exception of the three BUSCA radius measurements. This leads to an addition of an unquantifiable uncertainty into the radius measurements due to temporal effects\footnote{Such as stellar noise (e.g., un-occulted starspots) and different atmospheric observing conditions.}. However, the BUSCA radius measurements were collected simultaneously and therefore are not affected by temporal effects. Examining the BUSCA radius measurements, we can see a linear negative gradient ($\lambda \rightarrow \infty$ while $R_p \rightarrow 0$). This is indicative of a Rayleigh--like scattering slope. The theoretical transmission spectrum which was generated with the cloud modelling approach {\bf(ii)} used a bimodal cloud particle distribution to simulate a cloud pressure deck (0.001\,bar) combined with a joint Rayleigh haze (scaling factor 100). The base spectra using the A0 cloud model did not exhibit a behaviour equivalent to a Rayleigh--like scattering haze, due to large cloud particle sizes and added Fe droplets \citep[see][]{Molliere2017}. This explains how the two best fitting transmission spectra disagree below 550\,nm. The bimodal cloud particle distribution transmission spectrum gives a superior agreement to the BUSCA radius measurements\footnote{Note the agreement between the BUSCA radius measurements and the transmission spectra, was determined after the spectrum was fitted to the entirety of the radius measurements in this work. Including the data points from \citet{Gibson2013}.}. When taking into account that the BUSCA measured radii are free from temporal uncertainties, there is a greater likelihood that the bimodal cloud particle distribution transmission spectrum is the correct interpretation of the data.

To explore the bimodal cloud particle distribution solution further we looked into variations between the grey cloud deck pressure and the haze scaling factor. To see if it is possible to reproduce such an excellent agreement with the BUSCA radius measurements (e.g., $\leq$\,0.5-$\sigma$), whilst still achieving a good fit (e.g., $\chi^2_{\nu} \approx 1.4$) with the entirety of the radius measurements used in this work. Increasing the cloud deck pressure will result in different temperatures being probed during the transmission observation, because the atmospheric temperature is a function of pressure. Moreover, atomic and molecular lines will gain in importance in the spectrum when compared to the cloud opacity, making, e.g., the alkali Na and K lines very strong in the spectral range studied here. However, this can be circumvented by altering the metal enrichment: by rescaling the temperature structure to lower pressures for higher metal enrichment or higher pressures for lower metal enrichment \citep{Molliere2015}. 

\begin{figure} \includegraphics[width=0.48\textwidth,angle=0]{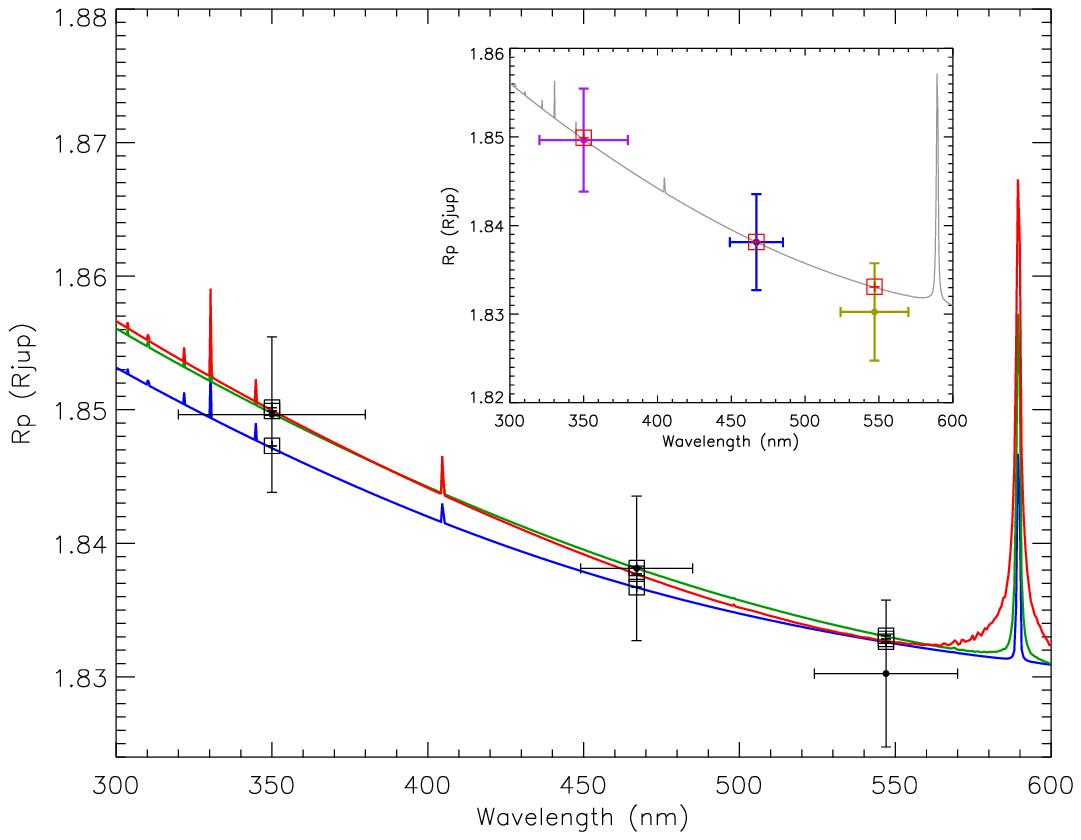}
\caption{\label{fig:three-bimodal-spectra}The three BUSCA planetary radius measurements compared to the three bimodal cloud particle distribution transmission spectra generated using {\tt petitCODE} \citep{Molliere2015, Molliere2017}. The transmission spectra are represented by the three solid lines: Red $\FeH = -0.45$, 0.01\,bar and 10$\times$ Rayleigh scaling. Green $\FeH = 0.55$, 0.001\,bar and 100$\times$ Rayleigh scaling. Blue $\FeH = 1.55$, 0.0001\,bar and 1000$\times$ Rayleigh scaling. The black open squares are the passband averages of the transmission spectra models, and are shown at the central wavelengths of the relevant passbands. The insert shows the comparison between the BUSCA radius measurements and the bimodal cloud particle distribution transmission spectrum with $\FeH = 0.55$, 0.001\,bar and 100$\times$ Rayleigh scaling.} \end{figure}

To show this, we took our clear base transmission spectra, one with an increase and the other with a decrease by an order of magnitude in the metal enrichment (e.g., $\FeH = -0.45$ and $\FeH = 1.55$). Cloud opacity was added to each of the transmission spectra. A cloud pressure deck of 0.01\,bar and a lower Rayleigh haze scaling of 10 were used as proxies for the cloud properties for the transmission spectrum with $\FeH=-0.45$. A cloud pressure deck of 0.0001\,bar and a higher Rayleigh haze scaling of 1000 were used for the transmission spectrum with $\FeH = 1.55$. The two transmission spectra were then fitted to the planetary radius measurements as described in Section\,\ref{Ray and cloud}. The best fit for the high and low pressure transmission spectra was found to be $\chi^2_{\nu} = 1.8$ and $\chi^2_{\nu} = 1.3$ respectively. The low pressure transmission spectrum with $\FeH = 1.55$ is shown in the lower right panel of Fig.\,\ref{fig:rp_lambda2}. The agreement between the two new bimodal cloud particle distribution transmission spectra and the BUSCA radius measurements was 0.21-$\sigma$ (0.01\,bar) and 0.37-$\sigma$ (0.0001\,bar). Fig.\,\ref{fig:three-bimodal-spectra} shows the comparison between the three bimodal transmission spectra and the three radius measurements from BUSCA. Fig.\,\ref{fig:three-bimodal-spectra} shows how it is not possible to discern the difference between the three bimodal cloud particle distributions, when using the BUSCA data alone. This can be explained due to a degeneracy between the cloud pressure deck and Rayleigh haze scaling factor. In the case of a much lower metal enrichment then the importance of the molecules with respect to Rayleigh scattering would go down. Therefore a lower value of the Rayleigh haze scaling factor would be needed to yield a strong Rayleigh scattering slope when compared to the molecular features. The BUSCA planetary radius measurements do though, constrain the ratio between the strength of the grey absorbing cloud deck relative to the haze component. 

The results from comparing the three bimodal cloud particle distribution transmission spectra with the planetary radius measurements used in this work show that varying the strengths of the cloud pressure deck and the Rayleigh haze scaling factor has little effect on the final fit or the agreement between the transmission spectra and with that of the BUSCA radius measurements. While in essence the three bimodal cloud particle distribution spectra are variations of the same solution, their unison good fit to the planetary radius measurements combined with their excellent agreement to the BUSCA radius measurements, adds considerable weight to a bimodal cloud particle distribution solution instead of a unimodal cloud particle distribution. A bimodal cloud particle distribution solution would also explain the discrepancies in the results found from previous studies which detected either a grey absorbing cloud deck or a Rayleigh--like scattering haze.

To study the structure of a bimodal cloud particle distribution we looked into two different scenarios: A Rayleigh--like haze stacked above a grey absorbing cloud deck\footnote{This is the bimodal cloud particle distribution transmission spectrum using a metal enrichment $\FeH = 0.55$, Rayleigh haze scaling factor $\times 100$ and 0.001\,bar cloud pressure (see Table\,\ref{tab:rp_lambda} and Figs.\ \ref{fig:rp_lambda2} and \ref{fig:three-bimodal-spectra}).}, or a Rayleigh--like haze and a grey absorbing cloud which are homogeneously extended throughout the atmosphere. To ascertain which scenario best agreed with the planetary radius measurements in this work we generated a new transmission spectrum using approach {\bf(ii)}. Both the cloud and haze particles were set at the same altitude to create a bimodal cloud particle distribution. In this scenario, at the shortest wavelengths the Rayleigh--like scattering would overpower the grey absorbing clouds due to its wavelength dependence ($\lambda^{-4}$) while at longer wavelengths the grey absorbing clouds overcome the Rayleigh--like scattering. To better understand this one can use a variation of Eq.\,1 from \citet{Lecavelier2008}.

\begin{equation}
R\left(\lambda\right) = R_0 + H \log \left[\kappa\left(\lambda\right)\right]
\end{equation}

\noindent where $R\left(\lambda\right)$ is the planetary radius at the evaluated wavelength $\left(\lambda\right)$, $R_0$ is the base planetary radius, $H$ is the atmospheric scale height and $\kappa\left(\lambda\right)$ is the wavelength dependent opacity.

In the two different scenarios $\kappa\left(\lambda\right)$ is determined differently. When the grey absorbing cloud particles are homogeneously extended throughout the atmosphere, $\kappa\left(\lambda\right)$ can be calculated as follows:

\begin{equation}\label{eq:kappa}
\kappa\left(\lambda\right) = \kappa_{R} \left(\frac{\lambda_0}{\lambda}\right)^4 + \kappa_{C}
\end{equation}

\noindent where $\kappa_R$ and $\kappa_C$ are the opacities for the Rayleigh haze and grey cloud deck respectively, $\lambda_0$ is a reference wavelength point, and $\lambda$ is the wavelength at which $\kappa\left(\lambda\right)$ is being evaluated. 

Eq.\,\ref{eq:kappa} shows that while $\kappa_C$ is constant at all wavelengths, $\kappa_R$ is dependent on wavelength. This allows a homogeneously extended Rayleigh--like haze and grey absorbing cloud to produce both a Rayleigh--like scattering slope at short wavelengths and a flat grey absorption at longer wavelengths. 

\begin{figure} \includegraphics[width=0.48\textwidth,angle=0]{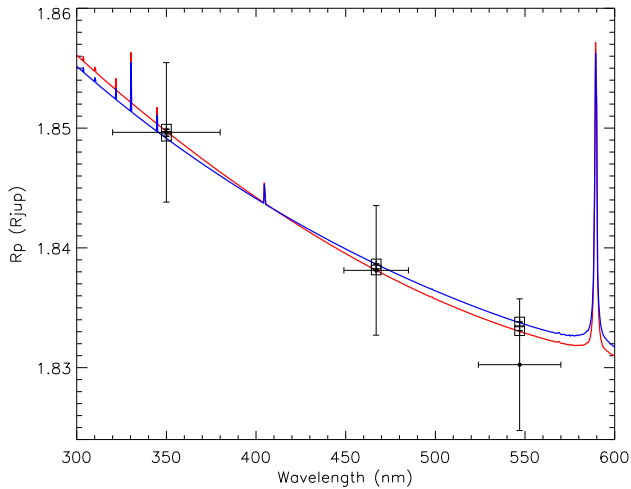}
\caption{\label{fig:rp_lambda-dual-model}The three BUSCA planetary radius measurements compared to two types of bimodal cloud particle distribution transmission spectra generated using {\tt petitCODE} \citep{Molliere2015, Molliere2017}. The red solid line represents the spectrum of the bimodal cloud particle distribution where the haze is stacked above the grey absorbing cloud deck. The blue solid line represents the spectrum of the bimodal cloud particle distribution where the grey cloud is homogeneously extended throughout the atmosphere. The black open squares are the passband averages of the transmission spectra models, and are shown at the central wavelengths of the relevant passbands.} \end{figure}

For the new transmission spectrum we introduced a grey absorber with an opacity of 0.01\,cm$^2$\,g$^{-1}$, instead of using a cloud pressure deck. The grey absorber was spread homogeneously throughout the entire atmosphere. We then fitted the transmission spectrum to the planetary radius measurements and obtained $\chi^2_{\nu} = 1.4$ for the best fit. This is in equal agreement to the Rayleigh--like haze stacked over a grey absorbing cloud deck (see Table\,\ref{tab:rp_lambda}). The new homogeneous grey absorber transmission spectrum agrees with the three BUSCA radius measurements at 0.27-$\sigma$. Fig.\,\ref{fig:rp_lambda-dual-model} shows the comparisons between the two bimodal cloud particle distributions; the stacked spectrum and the homogeneous grey absorber spectrum and with that of the three BUSCA radius measurements. From examining Fig.\,\ref{fig:rp_lambda-dual-model} it can be seen that with the current data at hand, it is not possible to discern between the two bimodal scenarios. We calculated that a 15-fold increase in the precision of the planetary radii measurements, would be needed to allow for a distinction between the two bimodal scenarios. Combined with this, both transmission spectra equally agree with the entirety of the radius measurements in this work. Therefore it is not possible to describe the cloud / haze properties any further, other than to mention the existence of a bimodal cloud particle distribution within the atmosphere of HAT-P-32\,A\,b. However, it should be noted that cloud particles large enough to generate a grey cloud opacity would tend to settle more strongly, potentially generating a thick cloud deck below the smaller haze particles, due to the smaller ratio between the surface area and mass of the larger cloud particles.


\section{Discussion and conclusions}
\label{Sec:w6conculsions}

HAT-P-32\,A\,b is an inflated, low density hot-Jupiter planet orbiting a hot ($T_{\rm eff} > 6250$\,K) host star, which makes the planet a perfect candidate to perform studies of planetary atmospheres. We used the collected data to determine the physical properties of the HAT-P-32 planetary system (Table\,\ref{tab:modelh32}) based on 11 new high-precision transit light curves and finding values which are consistent with those in the literature. We find the mass and radius of the host star to be $1.182\pm 0.041\Msun$ and $1.225\pm0.015\Rsun$, respectively. For the planet we find a mass of $0.80\pm 0.14\Mjup$, a radius of $1.807\pm0.022\Rjup$ and a density of $0.126\pm0.023\pjup$.

The 11 transits of HAT-P-32 were modelled using \prism\ and \gemc. This included two incomplete transits (observed in the Johnson V passband), for which we fixed, the sum of the fractional radii to the value determined by \citet{Gibson2013}. This is due to the sum of the fractional radii being directly related to the transit duration. It was set to the value from \citet{Gibson2013} to maintain homogeneity with the second analysis into the variations of $r_p$ with wavelength. This though, did not help with the partial transit observed on 2014/10/24. All of the parameter best fits from the partial transit observed on 2014/01/11 and the full transit observed in the Johnson V passband agree within their 1-$\sigma$ uncertainties. While, for the partial transit observed on the 2014/10/24 this is not the case for $k$, which disagrees with the other two Johnson V light curves by $\approx2\sigma$, however, the remaining parameters do agree within the 1-$\sigma$ uncertainties.

We observed one of the transits of HAT-P-32 using a simultaneous multi-band imaging instrument: BUSCA (using the Str\"{o}mgren \textit{u}, \textit{b} and \textit{y} passbands). Due to the known difficulty in attempting to obtain a high-precision light curve in the optical--UV via this technique, it was decided to select three passbands close to the UV-blue side of the optical spectrum. In conjunction with this we optimised the amount of defocusing used in the  Str\"{o}mgren \textit{b} passband to maximise the precision of the light curves in all three passbands. The resulting light curves produced are of decent quality ($\approx$1\,mmag scatter per point). In terms of the optical UV we were able to obtain a light curve with a rms scatter 1.08\,mmag per point. Our new u-band light curve has a threefold increase in precision compared to previous optical--UV light curves obtained using BUSCA \citep[i.e.,][]{Sou2015}. This has allowed an accurate measurement of the planetary radius in the optical--UV which is important for detecting either a Rayleigh scattering slope or the blue-edge detection of TiO in the planetary atmosphere.

Taking into account a blended M-dwarf companion (HAT-P-32\,B) within the defocused PSF of HAT-P-32\,A, combined with needing to study variations in the planetary radius with respect to wavelength, modifications were made to the modelling and optimisation codes \prism\ and \gemc. After making the new modifications, \prism\ and \gemc\ were then used in the analysis of the 11 transit in this work. The new versions are available from the first author.


\subsection{Rayleigh--like haze and grey absorbing cloud deck}
\label{Sec:Ray-dis}

The results from comparing the different planetary radii from each passband to the individual theoretical model atmospheres, confirms the results from \citet{Gibson2013, Mallonn2016b} by the detection of a grey absorbing cloud deck within the upper atmosphere of HAT-P-32\,A\,b. The planetary radius measurements from each passband were compared to the cloudy spectra, generated using a bimodal cloud particle distribution which consisted of a Rayleigh haze combined with a grey absorbing cloud pressure deck, showed the same level of agreement and confirms the possible presence of a Rayleigh haze within the atmosphere of HAT-P-32\,A\,b as found by \citet{Mallonn2016a}. However, the bimodal cloud particle distribution spectra give a superior agreement to the three radius measurements obtained from the BUSCA observations using the Str\"{o}mgren {\it u, b} and {\it y} passbands. Considering that the BUSCA radius measurements were the only observations obtained simultaneously in this study -- and therefore are not affected by unquantifiable temporal uncertainties -- allows for a greater weight to be placed on the accuracy of the set, compared to the remaining radius measurements  individually. A similar argument can be used to explain the discrepancy between \citet{Mallonn2016a} and \citet{Mallonn2016b}. \citet{Mallonn2016a} used transit spectroscopy where the observations were simultaneously obtained, while, \citet{Mallonn2016b} used an ensemble of photometric transit observations spanning from 2007 to 2016. 

At present there is supporting evidence for the presence of a Rayleigh scattering haze in the upper atmosphere of HAT-P-32\,A\,b from a prior investigation. \citet{Mallonn2016a} performed  transit spectroscopy to observe a transit of HAT-P-32\,A\,b. The data agreed with a Rayleigh scattering model in the optical--UV, with greater clarity below 550\,nm. However, a second study using photometric observations by \citet{Mallonn2016b} seemed to contradict the original results. \citet{Mallonn2016b} mentions a discrepancy between the two data sets in the redder wavelengths and attributed this to systematics in the measurement from \citet{Mallonn2016a}. The new data presented by \citet{Mallonn2016b} showed equal weight towards either a Rayleigh scattering model or that of a flat cloudy spectrum. When the data from both studies were combined and examined with a restricted wavelength range of $\lambda < 720$\,nm, the results showed a strong agreement with a flat cloudy spectrum. The Rayleigh scattering transmission spectrum which was generated by the model from \citet{Fortney2010} which was used in \citet{Mallonn2016a, Mallonn2016b} gives a continuous gradient from 350\,nm to 950\,nm. However, in other cases where a Rayleigh scattering slope has been found, it is generally detected below 550\,nm \citep[e.g., WASP-12\,b:][]{Sing2013}. When examining the combined data from both \citet{Mallonn2016a} and \citet{Mallonn2016b} it can be seen that the data could support a bimodal cloud particle distribution transmission spectrum approach, and so, support the detection of a Rayleigh--like scattering haze.

There have been previous exoplanet studies into the variation of the planetary transit depth at various wavelengths in the optical--UV photometric bands \citep[e.g.,][]{Turner2013,Turner2016}. In an attempt to measure the magnetic field of TrES-3\,b \citet{Turner2013}, obtained nine transits, four of which was observed using a Bessell\,U filter between June 2009 and April 2012. When examining the transit depths they determined a non-detectable change in the planetary radius between the Bessell\,U and Harris\,V filters. However, the major difference between the optical-UV transits in this work compared to the optical-UV transits from \citet{Turner2013}, is that for this work the three BUSCA transits were observed simultaneously and have an average rms scatter of $\approx$1\,mmag. While, the Bessell\,U transits from \citet{Turner2013} where each taken on different nights (introducing temporal uncertainties) and are of poorer quality, with a rms scatter of 4.1\,mmag for the four combined phased Besselll\,U transits. Therefore, masking any spectral features or changes between the near--UV bands.

\citet{Turner2016} recently completed an observational survey of 15 exoplanets, collecting photometric light curves in the optical-UV (Bessell-U, Harris-B, V and R). The overall results showed a non-detectable change in the planetary radius between the optical and near--UV, for 10 of the exoplanets, while, the other five exoplanets results; indicate the presence of an aerosol / Rayleigh scattering process. This survey supports the case that a scattering slope in the optical--UV is detectable using ground-based photometry.

\citet{Gibson2013} performed transit spectroscopy to observe the atmosphere of HAT-P-32\,A\,b\footnote{The results of their analysis are included in this work.} using GMOS on the Gemini North telescope. Through their analysis they determined that a grey absorber (cloud) was masking spectral features. A Rayleigh scattering slope is most prominent below 550\,nm, which lies beyond of the observing range of \citet{Gibson2013} (520--930\,nm). In this analysis it was found that the data from \citet{Gibson2013} agreed with the A0 cloudy \citep{Ackerman2001} transmission spectrum ($\chi_\nu^2 = 0.8$) and the bimodal cloud particle distribution spectrum ($\chi_\nu^2 = 0.9$). Therefore, we confirm that a grey absorber / cloud deck is present within the atmosphere of HAT-P-32\,A\,b.

The variations between the observed planetary radius with wavelength due to Rayleigh scattering is dependent on a power-law relation between wavelength and the mean cross-section of the atmospheric scattering particles \citep{Lecavelier2008}. The power-law coefficient which corresponds to Rayleigh scattering is $\alpha = -4$ \citep{Lecavelier2008}. To ascertain the validity of the Rayleigh scattering slope detection we fitted a straight line to the Str\"{o}mgren \textit{u}, \textit{b} and \textit{y} radii against wavelength and determined the gradient $\left(\frac{d R_b\left(\lambda\right)}{d ln \lambda} \right) $. We then calculate the temperature of the planet's terminator $T_{ter}$, due to the slope of the planetary transmission spectrum being proportional to $\alpha T_{ter}$:

\begin{equation}
\alpha T_{ter} = \frac{\mu g}{k_B} \frac{d R_b\left(\lambda\right)}{d ln \lambda}
\end{equation}

\noindent where $\mu$ is the mean molecular weight, which is taken as $2.3\,amu$, $g$ is the surface gravity and $k_B$ being the Boltzmann constant.

By assuming that the gradient is induced by Rayleigh scattering and therefore setting $\alpha=-4$, gives a temperature for the terminator of $1518\pm 345$\,K, which fits (within the 1-$\sigma$ uncertainty) with the equilibrium temperature, $T^{'}_{eq}\approx 1800$\,K (see Table\,\ref{tab:modelh32}) and is still below the day-side temperature of $\approx 2050$\,K \citep[see][]{Zhao2014}. Though it is not in agreement with the value found by \citet{Mallonn2016a} ($890\pm228$\,K), however, \citet{Mallonn2016a} states that their value appears to underestimate the true temperature of the terminator.

As a second validation check, the slope of the planetary transmission spectrum can be used to calculate the planetary mass and the result then compared to our previous result (see Table\,\ref{tab:modelh32}). For this we use the \textit{MassSpec} concept by \citet{Dewit2013}. Because the pressure scale height is dependent on the surface gravity and therefore the planetary mass, while, also being dependent on the power-law relation between wavelength and the mean cross-section of the atmospheric scattering particles, we can use our derived gradient to calculate the planetary mass using the following equation \citep[see][and references therein]{Sou2015b}:

\begin{equation}
M_b = -\frac{\alpha k_B T^{'}_{eq}\left[R_b\left(\lambda\right)\right]^2} {\mu G \frac{d R_b\left(\lambda\right)}{d ln \lambda}}
\end{equation}

\noindent where $G$ is the gravitational constant.

We applied \textit{MassSpec} to our determined gradient and by assuming $\alpha=-4$, we find a planetary mass of $0.94\pm0.16$\,\Mjup\ for HAT-P-32\,A\,b. This is in good agreement to the value found from analysing the transit light curves ($0.80\pm0.14$\,\Mjup).

Recently \citet{Batalha2017} has called in question the use of \textit{MassSpec} for determining the mass of small planets ($<3\,R_\oplus$), due to the uncertainty in different types of dominant atmospheres (e.g., CH$_4$ or CO$_2$), and so, it is not possible to assume a value for $\mu$. This therefore, makes the use of \textit{MassSpec} difficult for use in determining the mass of small planets, with a range of different molecular constituents. As HAT-P-32\,A\,b is a hot Jupiter planet then the concerns of using \textit{MassSpec} brought up by \citet{Batalha2017} have no implications in this work.

The results from this work show a possible detection of a bimodal cloud particle distribution ($\chi_\nu^2 = 1.4$) within an atmosphere comprising of a scaled solar metal enrichment $\left(\rm{Fe/H} = 0.55\right)$ and a solar C/O number ratio for HAT-P-32\,A\,b. The results from this work help towards confirming the tentative findings from \citet{Mallonn2016a}. The gradient of the Rayleigh--like haze scattering slope provides an extra validation by allowing the determination of the temperature at the terminator (1518$\pm$345\,K) and the planetary mass (0.94$\pm$0.16\,\Mjup), which, are confirmed through the separate analysis of the transit light curves. By obtaining high-precision ($\approx$\,1\,mmag) transit light curves in the optical-UV, this work shows that these measurements are important in the detection of either a Rayleigh scattering haze or a rich TiO atmosphere from ground based simultaneous multi-band photometry.


\section{Acknowledgements}
\label{sec:Acknow}

We would like to thank the anonymous referee for their helpful comments and in particular their comments regarding the descriptions of the theoretical transmission spectra used in this work. This paper incorporates observations collected at the Centro Astron\'{o}mico Hispano Alem\'{a}n (CAHA) at Calar Alto, Spain, operated jointly by the Max-Planck Institut f\"{u}r Astronomie and the Instituto de Astrof\'{i}sica de Andaluc\'{i}a (CSIC). Together with observations collected at the Observatorio Astron\'{o}mico Nacional at San Pedro M\'{a}rtir. JTR acknowledges financial support from  USRA (Universities Space Research Association), ORAU (Oak Ridge Associated Universities) and NASA in the form of a Post-Doctoral Programme (NPP) Fellowship. DR acknowledges financial support from the Spanish Ministry of Economy and Competitiveness (MINECO) under the 2011 Severo Ochoa Program MINECO SEV-2011-0187.

The following internet-based resources were used in the research for this paper: the NASA Astrophysics Data System; the SIMBAD database and VizieR catalogue access tool operated at CDS, Strasbourg, France; and the ar$\chi$iv scientific paper preprint service operated by Cornell University.


\end{document}